\documentclass[11pt]{article}
\pdfoutput=1

\usepackage{jheppub} 
\usepackage{amsmath,amssymb,epsfig,amsfonts}
\usepackage{epsf}
\usepackage{makeidx}
\usepackage{slashed}
\usepackage{verbatim} 
\usepackage{float}
\restylefloat{table}
\usepackage{pdflscape}
\usepackage{tikz}
\usepackage{pifont}
\usepackage{array}
\usepackage{youngtab}
\usepackage{multirow}
\usepackage{xcolor}
\usepackage{ulem}
\usepackage{cleveref}
\usepackage{tensor}

\makeatletter

\newcommand{\be}{\begin{equation}}
\newcommand{\ee}{\end{equation}}
\newcommand{\ba}{\begin{aligned}}
\newcommand{\ea}{\end{aligned}}

\newcommand{\GR}{\mathcal{G}^{R}}
\newcommand{\GI}{\mathcal{G}^{I}}

\newcommand\D{\mathcal{D}}

\newcommand{\dd}{\mathrm{d}}
\newcommand{\me}{\mathrm{e}}
\newcommand{\ii}{\mathrm{i}}

\newcommand{\vol}{\mathrm{vol}}

\newcommand{\Imag}{\mathrm{Im}\, }
\newcommand{\Real}{\mathrm{Re}\, }

\newlength{\sswidth}

\newcommand{\M}{\mathcal{M}}

\newcommand{\lb}{\left(}
\newcommand{\rb}{\right)}

\newcommand{\gads}{G_{\text{AdS}}}

\newcommand{\alphai}{\alpha}

\newcommand{\bea}{\begin{eqnarray}}
\newcommand{\eea}{\end{eqnarray}}

\newcommand{\R}{{\mathbb R}}
\newcommand{\Z}{{\mathbb Z}}

\def\Im{\mathop{\mathrm{Im}}\nolimits}

\def\Re{\mathop{\mathrm{Re}}\nolimits}

\def\unit{{1\kern-.65ex {\rm l}}}
\def\1{{1\kern-.65ex {\rm l}}}

\newcommand{\ads}{\text{AdS}}

\newcount\hour \newcount\minute
\hour=\time \divide \hour by 60
\minute=\time
\def\now{%
\ifnum \hour<13
  \ifnum \hour=0 \advance \hour by 12 \number\hour:\else \number\hour:\fi%
     \ifnum \minute<10 0\fi%
     \number\minute%
\ A.M.%
\else \advance \hour by -12 \number\hour:%
  \ifnum \minute<10 0\fi%
  \number\minute%
  \ P.M.%
\fi%
}

\makeatother

\begin{document}

\baselineskip=18pt  
\numberwithin{equation}{section}  
\allowdisplaybreaks  


%
%


\thispagestyle{empty}


\vspace*{1cm} 
\begin{center}
{\LARGE $\mathcal{N}=(0,2)$ AdS$_3$ Solutions of Type IIB and F-theory with Generic Fluxes} 
 \vspace*{1.8cm}
 
 \renewcommand{\thefootnote}{}
\begin{center}
 {Christopher Couzens
 \footnotetext{c.a.couzens@uu.nl}}
\end{center}
\vskip .2cm

 \vspace*{.5cm} 
{ Institute for Theoretical Physics and, \\
  Center for Extreme Matter and Emergent Phenomena, \\
  Utrecht University, Princetonplein 5, 3584 CE Utrecht, The Netherlands}\\
  
 {\tt {}}

\vspace*{0.8cm}
\end{center}

 \renewcommand{\thefootnote}{\arabic{footnote}}
 
\begin{center} {\bf Abstract } \end{center}

\noindent 
We consider the most general AdS$_3$ solutions of type IIB supergravity admitting a dynamical SU$(3)$ structure and preserving $\mathcal{N}=(0,2)$ supersymmetry. The analysis is broken into three distinct classes depending on whether the dynamical SU$(3)$ structure degenerates to a strict SU$(3)$ structure. The first class we consider allows for a holomorphically varying axio-dilaton consistent with the presence of $(p,q)$ 7-branes in addition to D3-branes and $(p,q)$ 5-branes. In the remaining two classes the axio-dilaton may vary but does not do so holomorphically. The second class of solution allows for 5-branes and 1-branes but no D3-branes, whilst in the final class all branes can be present. We illustrate our results with examples of such solutions including a new infinite family with all fluxes but the axion turned on. 



\newpage

\tableofcontents
\printindex


\section{Introduction}

There has been a recent resurgence in supersymmetric AdS$_3$ backgrounds in string theory. This is partly because two-dimensional CFTs provide an excellent testing ground for AdS/CFT since they are generally more tractable than their higher dimension cousins. Moreover, with the recent advent of c-extremization \cite{Benini:2012cz,Benini:2013cda} it has become possible to compute the central charges of strongly coupled two-dimensional IR fixed points using just UV data. The analogous extremization principle in gravity \cite{Couzens:2018wnk, Gauntlett:2018dpc,Gauntlett:2019pqg}\footnote{See \cite{Hosseini:2019use} for a proof of c-extremization and its geometric dual for toric theories.} has been understood for backgrounds with only five-form flux \cite{Kim:2005ez}. However, since c-extremization works independently of the brane construction of the 2d SCFT it is natural to conjecture that there are similar extremization principles for all AdS$_3$ backgrounds in string and M-theory preserving $\mathcal{N}=(0,2)$ supersymmetry. 

Progress in understanding the geometric dual of c-extremization was only made possible after the underlying geometry of the class of solutions consisting of just five-form flux was classified in \cite{Kim:2005ez} and further studied in \cite{Gauntlett:2007ts}. To make progress in extending the geometric dual of c-extremization to more general classes of AdS$_3$ solutions one must thus first derive the sufficient and necessary conditions for preserving supersymmetry \`a la \cite{Kim:2005ez}. This is the motivation behind this paper and we classify all supersymmetric AdS$_3$ solutions of type IIB supergravity preserving $\mathcal{N}=(0,2)$ with arbitrary flux configurations and an SU$(3)$ structure. 

Many works have been produced in classifying and studying AdS$_3$ solutions, whether in string theory or M-theory, yet a unified result for type IIB supergravity with arbitrary fluxes, $\mathcal{N}=(0,2)$ supersymmetry and a dynamical SU$(3)$ structure was until now lacking. Various aspects of type IIB supergravity with an AdS$_3$ factor have been discussed in \cite{Kim:2005ez, Donos:2008ug, Couzens:2017way, Eberhardt:2017uup, Couzens:2017nnr, Passias:2019rga}. Solutions preserving $\mathcal{N}=(0,2)$ have been considered in \cite{Kim:2005ez, Donos:2008ug, Couzens:2017nnr}. The first considers purely five-form flux configurations. Whilst the latter two include additional fluxes, they are not the most general configurations possible. In \cite{Couzens:2017way} $\mathcal{N}=(0,4)$ solutions with five-form flux and varying axio-dilaton were considered and \cite{Eberhardt:2017uup} considered $\mathcal{N}=(2,2)$ geometries with pure NS-NS flux. This work is an extension of all these results. Finally \cite{Passias:2019rga} considered $\mathcal{N}=(0,1)$ solutions with a particular choice of three-form flux. By making a suitable ansatz later, they find a class of $\mathcal{N}=(0,2)$ solutions generalizing \cite{Kim:2005ez, Donos:2008ug, Couzens:2017nnr}, this class will be recovered as a special case of our results. Solutions of these various classifications can also be found in \cite{Gauntlett:2006ns,Gauntlett:2006af,Benini:2015bwz,toappear}, see also \cite{Jeong:2014iva} for some TsT-dual solutions.

Important results in other supergravity theories for AdS$_3$ solutions have also been made. In massive type IIA $\mathcal{N}=(0,4)$ solutions have been classified and their field theories investigated in \cite{Lozano:2019ywa,Lozano:2019zvg,Lozano:2019emq,Lozano:2019jza, Lozano:2015bra}. Further work for type IIA can be found in \cite{Macpherson:2018mif, Dibitetto:2018ftj}. Results for Heterotic supergravity can be found in \cite{Beck:2015gqa}, whilst results in M-theory can be found in \cite{Martelli:2003ki,Colgain:2010wb}.

In this work we shall classify all $\mathcal{N}=(0,2)$ solutions of type IIB supergravity with arbitrary fluxes where the internal manifold admits an SU$(3)$ structure. We shall not make any assumptions on the form of the fluxes at any point in this work. The necessary and sufficient conditions for supersymmetric solutions are phrased in terms of torsion conditions for spinor bilinears after employing the G-structure formalism \cite{Gauntlett:2002sc}. In seven dimensions and with an SU$(3)$ structure imposed, this is determined by three differential conditions determining a one-form, a real two-form and a complex three-form. The vector dual to the one-form foliates the internal manifold and therefore we reduce the torsion conditions on to this base space. In general the base is complex but it need not be (conformally) K\"ahler. It does however satisfy the weaker (conformally) balanced condition.

The paper is organised as follows. In section \ref{Sec:susy} we derive the torsion conditions for an arbitrary G-structure. We specialise the G-structure to a dynamical SU$(3)$ structure in section \ref{Sec:SU(3)} and derive the necessary and sufficient conditions for a supersymmetric solution. The analysis must be split into three classes depending on the scalar bilinears which determine whether the dynamical SU$(3)$ structure reduces to a strict one. Some representative examples of solutions are given in section \ref{Sec:Sols} among which is a new infinite class of solutions. We conclude in section \ref{Sec:Conclude}. We have relegated some technical material and useful identities to two appendices.


\section{Conditions for supersymmetry in seven dimensions}\label{Sec:susy}

In this section we begin our journey of finding supersymmetric AdS$_3$ solutions of type IIB with generic fluxes. We construct the most general ansatz for the metric and fluxes preserving SO$(2,2)$ symmetry and reduce the ten-dimensional supersymmetry variations onto a seven-dimensional internal space $X_7$. We present the reduced supersymmetry equations that will form the starting point of our analysis.

By using G-structure techniques we will analyse the necessary and sufficient conditions for the preservation of supersymmetry by reformulating the supersymmetry equations, \eqref{PKSE}-\eqref{dKSE}, in terms of both algebraic and differential conditions on a set of differential forms constructed from spinor bilinears defined in table \ref{Spinor-bilinear table}. We spare the reader the tedious details of the derivation and present just the results. We have added a parameter $\alphai=\pm$ which determines the chirality of the preserved supersymmetry on the boundary of AdS. We have included the higher form torsion conditions for completeness despite the fact that they are typically implied by the lower form ones. 

Since we will specialise to the case of $\mathcal{N}=(0,2)$ supersymmetry, one expects the existence of an \emph{R-symmetry vector}. This is a Killing vector of the full solution both metric and fluxes and is holographically dual to the R-symmetry of a putative SCFT dual. We show that such a Killing vector always exists for solutions preserving $\mathcal{N}=(0,2)$ supersymmetry. We end the section by showing which equations of motion and Bianchi identities are implied by supersymmetry and which need to be imposed by hand.


\subsection{Reduced supersymmetry equations}

The focus of this work is to find $\mathcal{N}=(0,2)$ supersymmetric bosonic backgrouds of type IIB supergravity\footnote{We shall follow the conventions for type IIB supergravity as given in \cite{Gauntlett:2005ww}, to which we refer the reader for further details. In particular we will work in Einstein frame and use the formalism where the SU$(1,1)$ symmetry of type IIB is realised linearly.} fully preserving an $SO(2,2)$ symmetry. The ten-dimensional metric in Einstein frame is taken to be the warped product 
\be
\dd s^2_{10}= \me^{2 \Delta} \big( \dd s^{2}(\text{AdS}_3)+ \dd s^{2}(X_{7})\big)~,
\ee
where $\Delta \in \Omega^{0} (X_{7}, \mathbb{R})$. The metric on AdS$_3$ has been normalised such that $R_{ab}=-2 m^{2} g_{ab}$, with $m$ an arbitrary length scale related to the cosmological constant of AdS space. The remaining fluxes are expanded as
\begin{align}
F^{(5)}=(1+\star_{10}) F\wedge\dd \vol(\text{AdS}_3)~,~~ G^{(3)}= \gads \dd \vol(\text{AdS}_{3}) + G~,
\end{align}
with $F\in \Omega^{(2)}(X_{7},\mathbb{R})$, $\gads\in \Omega^{(0)}(X_{7},\mathbb{C})$ and $G\in \Omega^{(3)}(\mathbb{C})$. Moreover the axio-dilaton is taken to be a complex function of $X_{7}$ only. 
The reduction of the flux Bianchi identities and equations of motion with this ansatz are:
\begin{align}\label{reducedEOMS}
\dd F&= \Im[ \gads G^{*}]~,~~~\dd \star F=\frac{\ii}{2} G \wedge G^{*}~,\\
\mathcal{D} G&=-P \wedge G^{*}~,~~ \mathcal{D}\gads =-\gads^{*} P ~,~~ \mathcal{D}( \me^{4 \Delta} \star G)=\me^{4 \Delta}P \wedge \star G^{*} + \ii \gads \star F - \ii F \wedge G~.\nonumber
\end{align}

To proceed we must decompose the two ten-dimensional Majorana--Weyl Killing spinors, $\epsilon_{i}$ of type IIB supergravity under
\be
\text{Spin}(1,9) \rightarrow \text{Spin}(1,2) \times \text{Spin}(7)~.
\ee
We take 
\be\label{10dspinordecomp}
\epsilon_{i}= \psi \otimes \me^{\frac{\Delta}{2}} \chi_{i} \otimes \theta~,~~~\epsilon=\epsilon_{1}+\ii \epsilon_{2}\equiv \psi \otimes \me^{\frac{\Delta}{2}}\xi \otimes \theta~,
\ee
where $\chi_{i}$ are Spin$(7)$ Majorana spinors, $\theta$ is a two-component constant spinor satisfying $\sigma_{3}\theta=- \theta$ and $\psi$ is a Spin$(1,2)$ Majorana spinor on AdS$_3$ satisfying
\be\label{AdSKSE}
\nabla_{\mu} \psi= \frac{\alpha m}{2} \gamma_{\mu}\psi~.
\ee
The parameter $\alpha=\pm 1$ and corresponds to the chirality of the preserved supersymmetry on the boundary of AdS. For each choice of $\alpha$ there are two distinct Majorana Killing spinors satisfying \eqref{AdSKSE}. In general a solitary Dirac spinor $\xi$ will preserve $\mathcal{N}=(0,1)$ supersymmetry, however this is enhanced to $\mathcal{N}=(0,2)$ when the Killing spinor equation decouples $\xi$ from its charge conjugate in the reduced supersymmetry variations, see for example \cite{Donos:2008ug}. We make further comments about the G-structure and preserved supersymmetry in section \ref{Gstructure}.

With this spinor decomposition the ten-dimensional Killing spinor equations reduce to the following Killing spinor equations on $X_7$:
\begin{align}
0&=\gamma^{\mu}P_{\mu}\xi^{c}+\frac{\me^{-2 \Delta}}{4}\lb\slashed{G}-\ii \gads \rb \xi\label{PKSE}~,\\
0&=\lb \frac{1}{2} \partial_{\mu}\Delta\gamma^{\mu}-\frac{\ii \alphai m}{2} + \frac{\me^{-4 \Delta}}{8} \slashed{F} \rb \xi-\frac{\me^{-2\Delta}}{16}\lb 3\ii \gads+\slashed{G}\rb \xi^{c}~,\label{DeltaKSE}\\
0&=\lb\D_{\mu} +\frac{\ii \alphai m}{2}\gamma_{\mu} - \frac{\me^{-4\Delta}}{8} F_{\nu_1 \nu_2} \tensor{\gamma}{_{\mu}^{\nu_1 \nu_2}}\rb\xi+\frac{\me^{-2 \Delta}}{4}\lb \ii \gads \gamma_{\mu}+\frac{1}{2}G_{\mu\nu_{1}\nu_{2}}\gamma^{\nu_{1}\nu_{2}}\rb\xi^{c}~.\label{dKSE}
\end{align}
Simple manipulations of \eqref{PKSE} show that if $G$ vanishes then necessarily so does $\gads$. If this is the case then we fall within the class of geometries considered in \cite{Kim:2005ez, Couzens:2017nnr} and thus we shall restrict to $G\neq0$.



\subsection{Torsion conditions and general analysis}

In this section we give the torsion conditions for the spinor bilinears defined in table \ref{Spinor-bilinear table}. These are computed from the supersymmetry equations \eqref{PKSE}-\eqref{dKSE}.

\subsubsection*{Bilinear definitions}

We shall use the notation $\gamma_{(n)}$ to denote Clifford contraction. Table \ref{Spinor-bilinear table} defines all the possible spinor bilinears. Higher order forms are given by the Hodge star of the defined bilinears. 
\begin{table}[h]
\begin{center}
\begin{tabular}{|c|c|c|}

\hline
Scalars & $S_{ij}\equiv \bar{\xi}_{i} \xi_{j}$& $A_{ij}\equiv \bar{\xi}_{i}^{c} \xi_{j}$\\
\hline
One-forms& $K_{ij} \equiv  \bar{\xi}_{i} \gamma_{(1)} \xi_{j}$ & $B_{ij} \equiv \bar{\xi}_{i}^{c} \gamma_{(1)}\xi_{j}$\\
\hline 
Two-forms & $ U_{ij}\equiv \bar{\xi}_{i} \gamma_{(2)} \xi_{j}$ & $ V_{ij}\equiv \bar{\xi}_{i}^{c} \gamma_{(2)}\xi_{j}$\\
\hline
Three-forms & $X_{ij} \equiv \bar{\xi}_{i} \gamma_{(3)}\xi_{j}$ & $ Y_{ij}\equiv \bar{\xi}_{i}^{c} \gamma_{(3)} \xi_{j}$\\
\hline
\end{tabular}
\end{center}
\caption{Definition of the spinor bilinears}
\label{Spinor-bilinear table}
\end{table}
It is then simple, but slightly tedious, to compute the torsion conditions using \eqref{PKSE}-\eqref{dKSE} and some gamma matrix identities.\footnote{One could use the mathematica package \cite{Kuusela:2019iok} to check the conditions given here.}  

\vspace{5mm}

\subsubsection*{Scalar conditions}
\begin{align}
\dd \Real[S_{ij}]&=-\frac{m}{2}(\alphai_{i}-\alphai_{j})\Imag[K_{ij}]~,\label{ReSij}\\
\me^{-4 \Delta}\dd (\me^{4 \Delta}\Imag[S_{ij}])&=-\frac{3m}{2}(\alphai_{i}-\alphai_{j})\Real[K_{ij}]-\me^{-2 \Delta}\Real[\gads B_{ij}^{*}]+\me^{-4 \Delta} i_{\Im[K_{ij}]}F~,\label{ImSij}\\
\me^{-2 \Delta}\D(\me^{2 \Delta}A_{ij})&=-\frac{\ii m}{2}(\alphai_{i}-\alphai_{j})B_{ij}-A_{ij}^{*} P~.\label{A}
\end{align}


\subsubsection*{One-form conditions}
The one-form equations are\footnote{Our convention for contractions of $p$-forms into $q$-forms, ($p\leq q$) is
\begin{equation*}
i_{T^{(p)}}T^{(q)}= \frac{1}{p!}T^{(p)}_{\mu_{1}..\mu_{p}} \tensor{T}{^{(q)}^{\mu_{1}..\mu_{p}}_{\mu_{p+1}...\mu_{q}}}\dd x^{\mu_{p+1}}\wedge \dd x^{\mu_{q}}\frac{1}{(q-p)! }~.
\end{equation*}
We will also need notation for the contraction of a single index of a $p$-form into a $q$-form and their higher order generalisations. We shall define the notation $i^{(n)}_{T^{(p)}}T^{(q)}$ to mean contract the last $n$ indices of $T^{(p)}$ into the first $n$ indices of $T^{(q)}$, including the correct numerical factors.
}
\begin{align}
\me^{-4 \Delta}\dd \lb \me^{4 \Delta}\Real[K_{ij}]\rb =& m (\alphai_{i}+\alphai_{j})\Imag[U_{ij}] -\me^{-4\Delta} \Re[S_{ij}] F~,\label{ReK}\\
\me^{-4 \Delta} \dd \lb \me^{4 \Delta} \Im[K_{ij}]\rb =&- m(\alpha_{i}+\alpha_{j}) \Re[U_{ij}]-\me^{-4 \Delta} \Im[S_{ij}] F -\frac{\me^{-2 \Delta}}{2} \Re[ \gads^{*} V_{ij}]\nonumber\\
&-\frac{\me^{-2 \Delta}}{2} \Im[i_{B_{ij}} G^{*}]-\frac{\me^{-2 \Delta}}{2} \Re[ i_{V_{ij}} \star G^{*}]\label{ImK}
\\
\me^{-2 \Delta}\D(\me^{2 \Delta} B_{ij})=&P\wedge B_{ij}^{*}- \ii \me^{-2 \Delta} i_{\Im[K_{ij}]}G   \label{B}~.
\end{align}


\subsubsection*{Two-form conditions}
\begin{align}
\me^{-4 \Delta}\dd (\me^{4 \Delta}\Im[U_{ij}])=& -\frac{m}{2}(\alphai_{i}-\alphai_{j})\Re[X_{ij}]+\me^{-2 \Delta}\Im[A_{ij}^{*}G]~,\label{ImU}\\
\me^{-4 \Delta}\dd(\me^{4 \Delta}\Re[U_{ij}])=&\frac{m}{2}(\alphai_{i}-\alphai_{j})\Im[X_{ij}]+\frac{1}{2} \me^{-2 \Delta}\Im[i_{B_{ij}}\star G^{*}] +\frac{\me^{-2 \Delta}}{2} \Re[i^{(1)}_{V_{ij}}G^{*}]~,\label{ReU}\\
\me^{-6 \Delta}\D(\me^{6 \Delta}V_{ij})=&-\frac{3 \ii m}{2}(\alphai_{i}-\alphai_{j})Y_{ij}+\me^{-4\Delta} F \wedge B_{ij}+P\wedge V_{ij}^{*}\nonumber\\
&-\ii \me^{-2 \Delta}\gads \Re[X_{ij}]-\ii \me^{-2 \Delta}\Im[S_{ij}]G+\me^{-2\Delta}i_{\Im[K_{ij}]}\star G~.\label{V}
\end{align}


\subsubsection*{Three-form conditions}
\begin{align}
\me^{-8 \Delta}\dd(\me^{8 \Delta}\Im[X_{ij}])=&2m(\alphai_{i}+\alphai_{j})\star \Im[X_{ij}]-\me^{-4\Delta} F \wedge \Im[U_{ij}]\nonumber\\
&-\me^{-2 \Delta}\star \Im[\gads Y_{ij}^{*}]+\me^{-2 \Delta}\star \Re[A_{ij}G^{*}]~,\label{ImX}\\
\me^{-4 \Delta}\dd(\me^{4 \Delta}\Re[X_{ij}])=&-\me^{-4\Delta} i_{\Im[K_{ij}]}\star F+\me^{-2 \Delta}\Re[B_{ij}\wedge G^{*}]~, \label{ReX}\\
\me^{-6 \Delta}\D(\me ^{6 \Delta}Y_{ij})=& m (\alphai_{i}+\alphai_{j})\star Y_{ij}-\me^{-2 \Delta}G \wedge \Re[K_{ij}]+\ii \me^{-2 \Delta}\Re[S_{ij}]\star G-P \wedge Y_{ij}^{*}~.\label{Yij}
\end{align}


\subsubsection*{Four-form conditions}
\begin{align}
\me^{-8 \Delta}\dd(\me^{8\Delta}\star \Im[X_{ij}])=& -\frac{3m}{2}(\alphai_{i}-\alphai_{j}) \star \Re[U_{ij}]~,\label{Im*X}\\
\me^{-4 \Delta}\dd(\me^{4 \Delta} \star \Re[X_{ij}])=&-\frac{m}{2}(\alphai_{i}-\alphai_{j})\star \Im[U_{ij}]+ \me^{-4 \Delta} i^{(1)}_{F}\star  \Re[U_{ij}]\nonumber\\
&+\me^{-2 \Delta}\Im[G \wedge V_{ij}^{*}]+\me^{-2 \Delta}\star \Im[\gads V_{ij}^{*}] ~,\label{Re*X}\\
\me^{-6 \Delta}\D(\me ^{6 \Delta}\star Y_{ij})=&- \frac{\ii m}{2}(\alpha_i - \alpha_j) \star V_{ij}- P \wedge \star Y_{ij}^* - \me^{-4 \Delta} A_{ij} \star F   - \me^{-2 \Delta} G \wedge \Im[U_{ij}].\label{*Yij}
\end{align}


\subsubsection*{Five-form conditions}
\begin{align}
\me^{-4 \Delta}\dd (\me^{4 \Delta}\star \Im[U_{ij}])=&-m (\alphai_{i}+\alphai_{j})\star \Re[K_{ij}]+\me^{-2 \Delta}\Re[G\wedge Y_{ij}^{*}]+\me^{-4\Delta} F \wedge \star \Im[X_{ij}]~,\label{Im*Ured}\\
\me^{-8 \Delta}\dd (\me^{8 \Delta}\star \Re[U_{ij}])=& - m (\alphai_{i}+\alphai_{j})\star \Im[K_{ij}]~,\label{Re*Ured}\\
\me^{-6 \Delta}\D(\me^{6 \Delta}\star V_{ij})=& \me^{-2 \Delta}\gads \star \Im[K_{ij}]+ P \wedge\star  V_{ij}^{*}+\ii \me^{-2 \Delta} G \wedge \Re[X_{ij}] \nonumber \\
&+  \me^{-4 \Delta} i^{(1)}_{F}\star B_{ij}~.\label{*Vred}
\end{align}


\subsubsection*{Six-form conditions}
\begin{align}
\me^{-12 \Delta}\dd (\me^{12 \Delta}\star \Re[K_{ij}])&=\frac{5m}{2}(\alphai_{i}-\alphai_{j})\Im[S_{ij}]\dd \vol(X_{7}) + \me^{-2 \Delta} \Im[\gads^{*}A_{ij}] \dd \vol(X_{7}) \\
\me^{-8 \Delta}\dd(\me^{8 \Delta}\star \Im[K_{ij}])&=-\frac{m}{2}(\alphai_{i}-\alphai_{j})\Re[S_{ij}]\dd \vol(X_{7}) \label{Im*Kij} \\ 
\me^{-10 \Delta}\D(\me^{10 \Delta}\star B_{ij})&= -\frac{3 \ii m}{2} (\alpha_{i}-\alpha_{j}) A_{ij} \dd\vol(X_7)- P \wedge \star B_{ij}^{*}~.\label{*Bred}
\end{align}

\subsubsection*{Algebraic equations}

Moreover the differential conditions are supplemented with a number of algebraic conditions that can be derived from \eqref{PKSE} and \eqref{DeltaKSE}:
\begin{align}
P_{\mu}\Im[K_{ij}]^{\mu}&=0~, \label{PImK}\\
\partial_{\mu}\Delta \Im[K_{ij}]^{\mu}&=- \frac{m}{2}(\alphai_{i}-\alphai_{j})\Re[S_{ij}]~, \label{dDeltaImK}\\
2m(\alphai_{i}+\alphai_{j})\Im[S_{ij}]&=-\me^{-4\Delta}i_{\Re[U_{ij}]} F~, \label{S} \\
\me^{-2 \Delta}\Re[S_{ij}]\gads&=- m (\alphai_{i}+\alphai_{j})A_{ij}~,\label{GA}\\
i_{\Im[X_{ij}]} G&= \Re[S_{ij}] \gads~,\label{ImXintoG}\\
i_{B_{ij}^{*}} P&=\frac{\me^{-2 \Delta}}{4} ( i_{\Re[X_{ij}]}G+ \Im[S_{ij}] \gads)~,\label{BintoP}\\
4 \me^{2 \Delta}i_{\Re[K_{ij}]} P&=i_{Y_{ij}}G -\ii \gads A_{ij}~,\label{ReKintoP}\\
A_{ij}^{*}P -i_{P} V_{ij}^{*}&= -\frac{\me^{-2 \Delta}}{4}(i_{U_{ij}}G +\ii\, i_{X_{ij}}\star G -\ii \gads K_{ij})~,\label{APalg}\\
-\Re[S_{ij}] P - i_{P}\Im[U_{ij}]&=\frac{\ii \me^{-2 \Delta}}{4} i_{Y_{ij}}\star G~,\label{PGterm}\\
 \Im[S_{ij}] P - \, i_{P} \Re[U_{ij}]&= \frac{\me^{-2 \Delta}}{4}( i_{V_{ij}}G -\ii \gads B_{ij})~,\label{SPGterm}\\
\Im[S_{ij}] \dd \Delta-\frac{\me^{-4 \Delta}}{4}i_{\Im[K_{ij}]} F&= \frac{\me^{-2 \Delta}}{8}(-\Im[i_{V_{ij}}G^{*}]-6 \Re[B_{ij}^{*} \gads])~.\label{dDeltaalg}
\end{align} 

In the above torsion conditions we have allowed for the existence of more than one Killing spinor and therefore are preserving at least $\mathcal{N}=(0,2)$ supersymmetry. Instead if one wants to make contact with the $\mathcal{N}=(0,1)$ classification of \cite{Passias:2019rga} one should impose the existence of a single Dirac Killing spinor. With a little work, and by switching off $\gads$, one can derive the conditions presented there. Note that setting $\gads=0$ is equivalent to requiring $A_{ij}=0~,~ \forall i,j$.  If we decompose the Dirac spinor into the Majorana spinors $\chi_{i}$ in equation \eqref{10dspinordecomp} the condition that $A_{ij}=0$ is equivalent to the two Majorana spinors $\chi_{i}$ having equal norm $|\chi_{1}|=|\chi_{2}|$, which is indeed the additional condition imposed in \cite{Passias:2019rga}.


\subsection{Killing vector}\label{sec:Killing}

We shall show in this subsection that geometries admitting at least two Killing spinors preserving the same boundary chirality supercharges, necessarily have a universal Killing vector which we call the \emph{R-symmetry vector} in keeping with the literature. The R-symmetry vector is a symmetry of the full solution, not only of the metric. Recall that in the vanishing $G$ case only one Dirac Killing spinor was needed to define such a universal Killing vector, the inclusion of three-form flux gives an obstruction to this here. In principle, by a judicious choice of fluxes, such as in \cite{Donos:2008ug}, one could make the vector field $K_{ii}$ a Killing vector. This fine tuning will not lend itself to a general analysis and therefore we shall not analyse this choice of flux further. 

The supersymmetry conditions imply
\begin{align}
\nabla_{(\mu} \Im[K_{ij}]_{\nu)} = \frac{ m}{2} (\alpha_1 - \alpha_2) g_{\mu \nu} \Re[S_{ij}]~,
\end{align}
and therefore for $\alpha_1 = \alpha_2$, (i.e. $\mathcal{N}=(0,2)$) the vector dual to the one-form $\Im[K_{12}]$ is a Killing vector.  

 We have thus shown that there always exists a Killing vector generating a symmetry of the metric for any $\mathcal{N}=(0,2)$ preserving G-structure. To substantiate our claim that this Killing vector is dual to the R-symmetry of the field theory we must now show that this Killing vector is a symmetry of the full solution, i.e.
\begin{align}
\mathcal{L}_{\Im[K_{ij}]} F = \mathcal{L}_{\Im[K_{ij}]} G = \mathcal{L}_{\Im[K_{ij}]} \gads = \mathcal{L}_{\Im[K_{ij}]} P = \mathcal{L}_{\Im[K_{ij}]} \Delta  = 0~.
\end{align} 

The algebraic conditions \eqref{PImK} and \eqref{dDeltaImK} imply that $i_{\Im[K_{ij}]} P = 0$ and $i_{\Im[K_{ij}]} \dd \Delta=0$, from which it follows immediately that both $\Delta$ and $\tau$ have vanishing Lie derivatives along the R-symmetry vector. 

Taking the gauged derivative of equation \eqref{B} and using the Bianchi identity for $G$ implies that $\mathcal{L}_{\Im[K_{ij}]} G=0$. Finally taking the exterior derivative of \eqref{S} and using the Bianchi dientity for $F$ implies $\mathcal{L}_{\Im[K_{ij}]} F=0$.

We conclude that $\Im[K_{ij}]$ defines an R-symmetry vector preserving the full solution provided the Bianchi identities for $F$ and $G$ hold. 


\subsection{Bianchi identities and equations of motion}

We must check which equations of motion and Bianchi identities are automatically implied by supersymmetry. In the case of trivial three-form fluxes all but the equation of motion for the two-form $F$ was implied by supersymmetry. Instead the equation of motion for $F$ needed to be imposed separately and is equivalent to imposing the \emph{master equation} \cite{Kim:2005ez,Donos:2008ug, Couzens:2017nnr}. As in the aforementioned cases the Einstein equation and equation of motion for the axio-dilaton are implied by supersymmetry here. Only the $F$ and $G$ flux Bianchi identities and equations of motion need further consideration. We shall see that there are two cases that we need to consider depending on whether $\gads$ is vanishing or not. In the generic case when $\gads\neq0$ we find that all the equations of motion and Bianchi identities are imposed provided that the Bianchi identity for $G$ is satisfied. In particular the equation of motion for $F$ is automatically satisfied. 
Instead when $\gads=0$ we find that imposing the Bianchi identity for $G$ is not sufficient and one must also impose the equation of motion for the two-form $F$ which amounts to imposing the master equation derived in \cite{Passias:2019rga} or one of the less general ones in \cite{Kim:2005ez,Donos:2008ug,Couzens:2017nnr} depending on the choice of fluxes.

Integrability of the torsion conditions implies the following relations between the equations of motion and Bianchi identities:
\begin{align}
A_{ij} F_{\text{EOM}}&=-\me^{2 \Delta} G_{\text{Bianchi}}\wedge \Im[U_{ij}]~,\\
\ii \Re[S_{ij}] G_{\text{EOM}}&=\me^{4 \Delta} G_{\text{Bianchi}} \wedge \Re[K_{ij}]~.
\end{align}
In addition the Bianchi identity for $F$ \eqref{reducedEOMS} is imposed by \eqref{ReK} and \eqref{ImU} whilst the Bianchi identity for $\gads$ is imposed by \eqref{A} and \eqref{GA}. 

The result of this analysis shows that imposing the $G$ Bianchi identity implies that the $G$ equation of motion is always satisfied and in addition when $\gads\neq 0$ the equation of motion for $F$ is also satisfied. On the other hand when $\gads=0$ we find that one must impose the equation of motion for $F$ in addition, which as we shall see later, is equivalent to the condition of section 5 in \cite{Passias:2019rga}.


\subsection{Some simplifying observations for chiral theories}\label{scalars}

Before we end this section let us make a few remarks about the scalars in the solution when we consider chiral theories, i.e. $\alphai_{i}=\alphai$. It is easy to see from \eqref{ReSij} that $\Re[S_{ij}]$ is a constant for any choice of $i,j$. By a suitable constant rescaling of the spinor we may normalise the spinors such that the norms $S_{ii}$ (no sum) are 1. Furthermore, since $\Re[S_{12}]$ is a constant, we may  set $\Re[S_{12}]=0$ by a constant $GL(2,\R)$ rotation of the spinors, further details may be found in \cite{Couzens:2017way}. Therefore $\Re[S_{ij}]=\delta_{ij}$.

Inserting this into \eqref{GA} implies that  $A_{12}=0$ and $A_{11}=A_{22}\equiv A=-\frac{\alpha\me^{-2 \Delta}}{2m} \gads$. We have introduced the shorthand $A$ for this complex scalar related to $\gads$. 
Furthermore notice that for a constant axio-dilaton $\gads$ is also necessarily constant. The final non-trivial scalar is the real scalar $\Im[S_{12}]\equiv S$.


\section{Specifying a G-structure: the SU$(3)$ case}\label{Sec:SU(3)}

Until now our analysis has been completely general and we have not imposed a G-structure on the internal manifold.
We deviate from this general analysis in the remainder of the paper and in the this section we will impose a particular G-structure on the internal manifold.

\subsection{G-structure and supersymmetry preservation}\label{Gstructure}

We restrict our attention to solutions preserving $\mathcal{N}=(0,2)$ since we want the internal manifold to admit an R-symmetry vector, and for the dual field theory to admit an R-symmetry. As we remarked in the introduction, this is desirable since it allows for the use of c-extremization \cite{Benini:2012cz, Benini:2013cda} in computing the central charge and R-charges of the dual 2d SCFT. Without a continuous R-symmetry this is not possible. This is reflected in gravity by the fact that in the geometric dual of c-extremization \cite{Couzens:2018wnk, Gauntlett:2018dpc}\footnote{See \cite{Gauntlett:2019roi} for the geometric dual of $\mathcal{I}$-extremization and \cite{Hosseini:2019ddy,Kim:2019umc} for further developments.}, the existence of an R-symmetry vector plays a central role. Recall that the class of gravity solutions studied there had only five-form flux turned on, preserved $\mathcal{N}=(0,2)$ and admitted an SU$(3)$ structure. We want to to mimic the properties of the classification in \cite{Kim:2005ez} as much as possible whilst turning on arbitrary three-form fluxes and axio-dilaton. As such, we impose that the internal manifold admits two independent Killing spinors satisfying \eqref{PKSE}-\eqref{dKSE}, thereby preserving $\mathcal{N}=(0,2)$ supersymmetry\footnote{A key difference of turning on three-form flux to the case without is the counting of preserved supersymmetries by a single Dirac spinor. As is clear from the explicit expressions \eqref{PKSE}-\eqref{dKSE}, the supersymmetry equations without three-form flux are invariant under the multiplication of the Killing spinor by an arbitrary complex constant. By this mechanism one can trivially generate an additional real supercharge on the boundary. However as soon as one turns on a generic three-form flux this argument is no longer applicable and a single Killing spinor thus gives a single real supercharge, i.e. $\mathcal{N}=(0,1)$. This enhancement can be seen in \cite{Donos:2008ug} with the inclusion of transgression terms and in section 5 of \cite{Passias:2019rga}, the latter of which studies $\mathcal{N}=(0,1)$ solutions.}. With these two spinors we have a richer choice of G-structure than in \cite{Kim:2005ez}, however we shall focus on the SU$(3)$ G-structure case as was done in \cite{Kim:2005ez,Donos:2008ug,Couzens:2017nnr}. We have presented some details about SU$(3)$ structures in appendix \ref{app:SU(3)} and present the analysis of the SU$(3)$ G-structure in the following section. 

The static SU$(3)$ structure for solutions preserving $\mathcal{N}=(0,1)$ has recently been investigated in the case where $\gads=0$ but otherwise general $G$ flux in \cite{Passias:2019rga}. The authors of \cite{Passias:2019rga} also made a further ansatz in a later section thereby enhancing the amount of supersymmetry preserved to $\mathcal{N}=(0,2)$ and generalizing the results of \cite{Couzens:2017nnr} to include a particular three-form and the results of \cite{Donos:2008ug} to include a non-trivial axio-dilaton. We shall recover this as a special case. 

An SU$(3)$ structure in seven dimensions is specified by the existence of a single real vector defining a transverse foliation over a six-dimensional base admitting a canonical SU$(3)$ structure. This constitutes a real two-form $J$ and a $(3,0)$-form $\Omega$, which satisfy the algebraic conditions
\be
\frac{1}{3!} J^{3}= \dd \vol_{6}~,~~ \Omega \wedge \bar{\Omega}= -8 \ii~ \dd \vol_{6}~,~~J \wedge \Omega=0 ~.
\ee
Since we have two spinors forming this SU$(3)$ structure it is necessary to impose a condition on the spinor bilinear scalars. As we explain in appendix \ref{app:SU(3)} one must impose a relation between the scalar spinor bilinears $A\equiv A_{11}=A_{22}$ and $S\equiv \Im[S_{12}]$, namely
\be\label{SU(3)scalarcond}
|A|^{2} + S^{2}=1~.
\ee
One can then rewrite the two scalars as
\be
A=\me^{\ii \varphi} \sin \Theta~,~~ S= \cos\Theta~,
\ee
with $\Theta$ and $\varphi$ real functions. This parametrisation will not be used extensively in the computations, but should be kept in mind. There are two special limits to consider where $A$ or $S$ vanish (but not both!) and they require a separate investigation.

For clarity of the reduced torsion conditions we shall show how the classic $\ads_{3} \times S^3 \times S^3 \times S^1$ with only three-form fluxes fits into the classification and we shall present a new solution which fits into the more general class of solutions. This new solution is obtained by performing a beta deformation \cite{Lunin:2005jy} on the solution given in \cite{Gauntlett:2006ns} with a product $T^2$. Such a solution then admits both three-form flux, five-form flux and dilaton and possesses a parameter interpolating between vanishing and non-vanishing three-form flux. To the best of our knowledge such a solution has not appeared in the literature previously.

\subsection{Reduced torsion conditions}\label{sec:Reduced}

We shall now assume throughout the remainder of the paper that the internal manifold admits an SU$(3)$ structure.\footnote{One should note that the solution discussed in appendix F of \cite{Couzens:2017way} admits an identity structure. Of course given an identity structure one can still define an SU$(3)$ structure as considered here, however it would still be interesting to look into the identity structure case in the future.} Having imposed the structure we can simplify the above torsion conditions in terms of the invariant forms of the SU$(3)$ structure. It is easy to check, either through Fierz identities or by constructing an orthonormal frame, that the bilinears satisfy\footnote{For simplicity of notation we have defined
\begin{equation*}
K\equiv\Im[K_{12}]~,~~A\equiv A_{11}~,~~ S\equiv \Im[S_{12}]~.
\end{equation*}}
\begin{align}
\Im[K_{12}]&\equiv K = -e^7~,~~K_{11}=K_{22}= S K~,~~ B_{12}= \ii A K~,~~ \\
U_{12}&=J~,~~U_{11}=U_{22}= -\ii S J~,~~ V_{12}=A J~,~~\\
X_{11}&=-\ii K \wedge J -\ii |A| \Re[\Omega]~,~~X_{22}=-\ii K \wedge J +\ii |A| \Re[\Omega]~,~~X_{12}= S J \wedge K -\ii |A| \Im[\Omega]~,\\
Y_{11}&=-\ii A K \wedge J - \ii \frac{A}{|A|} (\Re[\Omega]-\ii S \Im[\Omega])~,~~Y_{22}=-\ii A K \wedge J + \ii \frac{A}{|A|} (\Re[\Omega]-\ii S \Im[\Omega])~,\nonumber\\
Y_{12}&=\frac{A}{|A|} (S \Re[\Omega]-\ii \Im[\Omega])~.\label{frame:Yij}
\end{align}
Here $e^7$ is a vielbein transverse to the six-dimensional canonical SU$(3)$ structure foliation. Note that the Killing vector has unit norm and since it defines a transverse foliation of the space we can introduce a local coordinate via
\be\label{Kcomp}
K^{\#}=- 2m \frac{\partial}{\partial \psi}~,~~~ K =- \frac{1}{2m}(\dd \psi + \sigma)~.
\ee
The metric then takes the form
\be
m^2\dd s^2(X_{7})=\frac{1}{4} (\dd \psi+ \sigma)^2 + \dd s^{2} (\mathcal{M}_{6})~,
\ee
and we reduce the torsion conditions onto the base of the $U(1)$ fibration $\M_{6}$.
In the remainder of this section we shall first reduce the torsion conditions to a minimal set without making any assumptions on the scalar bilinears. In later sections we shall consider the three cases separately.


\subsubsection*{Scalar conditions}

Let us first study the conditions arising from the scalar bilinears. The only non-redundant condition is obtained from equation \eqref{A} which implies the Bianchi identity for $\gads$. The condition following from \eqref{ImSij} is implied by later conditions.


\subsubsection*{One-form conditions}

Since $K_{11}=K_{22}$, we may use the torsion condition for $\Re[K_{ii}]$ to define the two-form $F$, 
\be\label{Fdef}
F=-2m \me^{4 \Delta}S J-\dd (\me^{4 \Delta} S K)~.
\ee
This takes a similar form to the two-form appearing in \cite{Kim:2005ez} and the later generalisations, differing only by the inclusion of the scalar $S$. It is clear that for $S=0$, which is one of the special cases we shall consider later, that the five-form flux is switched off. It is useful to decompose $F$ into a part with a leg along the Killing direction and a term without, we have
\be
F= K \wedge \dd (\me^{4 \Delta} S)+ \hat{F}~,~~~~ \hat{F}= -\me^{4 \Delta} S (2m J + \dd K)~.
\ee
Equation \eqref{B} implies
\be
i_{K} G=- \me^{2 \Delta} A \dd K
\ee
and therefore we may write $G$ as
\be\label{Gform}
G =- \me^{2 \Delta} A K \wedge \dd K + \hat{G}~,~~~~ i_{K} \hat{G}=0~.
\ee
The complex three-form $\hat{G}$ is as yet undetermined.

The final non-trivial one-form equation implies two conditions
\begin{align}
\me^{-8 \Delta} \dd (\me^{8 \Delta} |A|^2)&= \me^{-2 \Delta} i_{J} \star_{6} \Re[A \hat{G}^{*}]~,\label{d8DeltaAnorm}\\
0&= |A|^{2} \big(2 m J + \dd K +*_{6}( J\wedge \dd K )\big)~.\label{AdK}
\end{align}
We shall see later that the bracketed part of the second equation is in fact true even for $A=0$ and implies $i_{J}\dd K=-2m$.


\subsubsection*{Two-form conditions}

Equation \eqref{ImU} implies 
\be\label{dSJ}
\dd (\me^{4 \Delta} S J)= \me^{2 \Delta} \Im[A \hat{G}^{*}] =\frac{1}{2m} \Im[\gads^{*} \hat{G}]~.
\ee
This guarantees that the three-form $\Im[\gads^{*} \hat{G}]$ is closed as is necessary for the Bianchi identity of $F$ to be consistent. 
Equation \eqref{ReU} implies the two conditions
\begin{align}
|A|^{2} i^{(1)}_{J} \dd K &=0~,\\
\me^{-4 \Delta} \dd (\me^{4 \Delta} J)&= \frac{\me^{-2 \Delta}}{2} \big(*_{6} \Re[ A \hat{G}^{*}]+ i^{(1)}_{J} \Re[A \hat{G}^{*}]\big)~.\label{dJ}
\end{align}
Finally equation \eqref{V} implies
\be\label{AdJ}
A \me^{-4 \Delta} \dd (\me^{4 \Delta} J ) =2 A^{*} P \wedge J + \me^{-2 \Delta} ( *_{6} \hat{G} - \ii S \hat{G})~.
\ee
Depending on the case under consideration imposing that these three conditions are consistent for $J$ will impose conditions on the three-form flux $\hat{G}$. For the special case of $A=0$ it is easy to see that $J$ is conformally closed, and $\hat{G}$ satisfies a self-duality constraint. Instead in the $S=0$ case we see that $\Im[A \hat{G}^{*}]$ vanishes. We shall analyse these cases in further detail later. 


\subsubsection*{Three-form conditions}

The reduced three-form equations will be used to compute the derivative of the holomorphic three-form $\Omega$. From equation \eqref{ReX} we find
\be\label{JKcond}
S\big(2 m J + \dd K + *_{6}(J \wedge \dd K)\big)=0
\ee
and since both $A$ and $S$ cannot be simultaneously vanishing we have that the bracketed expression must vanish identically after using \eqref{AdK}.

Reducing equation \eqref{ImX} for the difference $X_{11}-X_{22}$ and for $X_{12}$ implies
\be\label{AOmega}
\me^{-8 \Delta} \dd ( \me^{8 \Delta} |A| \Omega)= -2  \ii m |A| K \wedge\Omega ~.
\ee
Therefore the internal manifold is complex when $A\neq0$ and the almost complex structure provided by $J$ is integrable. In fact we shall see shortly that the base $\M_{6}$ is complex in all cases, and as such we shall use that we can split $n$-forms into forms of bidegree $(p,q)$, $p+q=n$. Computing \eqref{ImX} for the sum $X_{11}+X_{22}$ one finds
\be
*_{6}\Re[A \hat{G}^{*}]= J \wedge *_{6} (J \wedge \Re[A \hat{G}^{*}])+ i_{J}^{(1)} \Re[A \hat{G}^{*}]~,
\ee
which implies that $\Re[A \hat{G}^{*}]$ has no $(3,0)$ nor $(0,3)$ part (this is implied by the algebraic conditions also), whilst it is satisfied identically for generic $(2,1)$ and $(1,2)$ three-forms. 

We obtain two further conditions on the holomorphic three-form $\Omega$ by reducing \eqref{Yij} for $Y_{11}-Y_{22}$ and $Y_{12}$ and taking suitable combinations:
\begin{align}
\me^{-6 \Delta} \mathcal{D} \lb \me^{6 \Delta} (1+S)\frac{A}{|A|}  \bar{\Omega}\rb &=2 \ii m (1+S) \frac{A}{|A|} K\wedge \bar{\Omega}+ (1-S)\frac{A^{*}}{|A|} P \wedge \bar{\Omega}~,\label{dAOmegab}\\
\me^{-6 \Delta} \mathcal{D} \lb \me^{6 \Delta} (1-S)\frac{A}{|A|} \Omega\rb &=-2 \ii m (1-S) \frac{A}{|A|} K \wedge \Omega +(1+S) \frac{A^{*}}{|A|}P \wedge \Omega\label{dAOmega}~.
\end{align}
The equation for the sum $Y_{11}+Y_{22}$ is trivial in this case. 
Note that the three equations for $\Omega$ given above all imply that the transverse foliation is a complex space.

\subsubsection*{Four-form conditions}

The only remaining torsion condition to impose is the four-form equation \eqref{Im*X} for the sum $X_{11}+X_{22}$ which implies
\be\label{balanced}
\me^{-8\Delta} \dd (\me^{8\Delta} J \wedge J)=0~.
\ee
This is the condition for a manifold to be conformally balanced. 

All other differential torsion conditions are implied by the conditions above and so this is a set of necessary conditions to impose. In the following sections we shall simplify the above conditions, however this must be done on a case by case basis depending on whether the SU$(3)$ structure is dynamical or strict. We first consider the case when $A=0$, before turning our attention to $A\neq0$. For $S=0$ the conditions simplify further and we present the $S=0$ case in the final part of this section.


\subsection{Special case 1: $A=0$}

Recall from \eqref{GA} that $A=0$ implies $\gads=0$ and therefore the geometry does not support the inclusion of D1-branes nor fundamental strings. The SU$(3)$ scalar condition \eqref{SU(3)scalarcond} implies $S=\pm1$. Without loss of generality we choose to set $S=1$, however for a better exposition it is useful to keep $S$ explicitly in places and therefore we employ a somewhat laissez-faire approach towards setting $S=1$.\footnote{The sign is correlated with the sign of the complex structure that we choose and is thus irrelevant. We have chosen $S=1$ so that the axio-dilaton is holomorphic as is the usual convention in F-theory.} Since the only solutions in this class are those contained in section 5 of \cite{Passias:2019rga} we shall be somewhat brief in this section and sketch only the most pertinent details.

Note that setting $A=0$ in the frame given in \eqref{frame:Yij} is a bit subtle. The term $\frac{A}{|A|}$ in the limit of $A=0$ simply becomes a phase which may be absorbed into the definition of $\Omega$. This subtlety does not extend to any other terms where the naive limit works and should be taken.

With the frame in hand let us reduce the necessary conditions derived in the previous section \ref{sec:Reduced}. The two-form flux $F$ is given by \eqref{Fdef}. Further it is trivial to see from \eqref{Gform} that $G$ has no leg on the Killing vector and is therefore defined only on the transverse foliation $\mathcal{M}_{6}$. From \eqref{dSJ} and \eqref{dJ} we see that the transverse foliation $\M_{6}$ admits a conformally closed two-form which will give rise to the K\"ahler form
\be
\dd (\me^{4 \Delta}J)=0~.
\ee
This clearly also satisfies the conformally balanced manifold condition \eqref{balanced}.
Moreover \eqref{AdJ} implies
\be
\star_{6} \hat{G}=\ii S \hat{G}~.
\ee
The sign of $S$ then fixes whether $\hat{G}$ is self-dual or anti self-dual. We absorb the conformal factor into the definition of the two-form by defining the new SU$(3)$ structure forms
\be
j= m^2 \me^{4 \Delta}J~,~~~\omega=m^3 \me^{6 \Delta}\Omega~.
\ee
The metric takes the form
\be
m^2 \dd s^2 =\frac{1}{4}(\dd \psi+\sigma)^2 + \me^{-4 \Delta} \dd s^{2}(X_{6})
\ee
with $X_{6}$ K\"ahler. 

Finally \eqref{dAOmegab} and \eqref{dAOmega} imply\footnote{We have set $S=1$ here, setting $S=-1$ will exchange $\omega \leftrightarrow \bar{\omega}$  below and holomorphic $\leftrightarrow$ anti-holomorphic.} 
\be
P \wedge \omega=0~,~~~  \D  \bar{\omega}= -\ii (\dd \psi +\sigma) \wedge \bar{\omega}~. 
\ee
The first condition shows that $\tau$ is a holomorphic function, this can also be derived from the algebraic condition \eqref{SPGterm}.
The second condition is precisely the result from \cite{Couzens:2017nnr} and gives the Ricci-form of the base $\rho$ in terms of the $\tau$ dependent connection $Q$ and the connection on the R-symmetry vector $\sigma$,
\be
\rho=\dd(\sigma-Q)~.
\ee
From \eqref{JKcond} and from the K\"ahler geometry identity
\be\label{KahleridentityR}
R_{\mu\nu}=-\tensor{J}{_{\mu}^{\tau}}\rho_{\tau \nu}~,
\ee
we find that the Ricci scalar is given by
\be
R=2|P|^2 + 8 \me^{-4 \Delta}~, 
\ee
exactly as in \cite{Couzens:2017nnr}.

Finally the algebraic conditions imply that $\hat{G}$ is a primitive\footnote{A $p$-form $u^{(p)}$, $p=\{2,3\}$ is primitive with respect to the complex hermitian form $J$ if $i_{J} u^{(p)}$=0.} $(2,1)$-form ($(1,2)$-form if one sets $S=-1$), and its Bianchi identity implies that it takes the form
\be
\hat{G}=\frac{\ii}{\sqrt{\tau_{2}}} (\tau \dd B - \dd C^{2}) \in \Omega^{(2,1)}(\M_6)~.
\ee 

The final condition to impose is the equation of motion for $F$, which is not imposed by supersymmetry in this case. It is trivial to see by using an amalgamation of the results in \cite{Donos:2008ug} and \cite{Couzens:2017way} that this implies precisely the master equation presented in \cite{Passias:2019rga}
\be
\square (R-2|P|^2) -\frac{1}{2} R^2+ R_{\mu\nu} R^{\mu\nu} +2 |P|^2 R -4 R_{\mu\nu} P^{\mu} P^{*\nu}-4 |G|^2=0~.
\ee
We have recovered the class of $\mathcal{N}=(0,2)$ solutions considered in \cite{Passias:2019rga} which they obtained upon imposing an ansatz in their $\mathcal{N}=(0,1)$ results. This gives a generalization of the results in \cite{Kim:2005ez,Donos:2008ug,Couzens:2017nnr} to include additional fluxes. As a byproduct of our analysis we have shown that this class of solutions is the unique one preserving $\mathcal{N}=(0,2)$, $\gads=0$ and an SU$(3)$ structure. 


\subsection{$A\neq 0$ case}

In the following we will analyse the class of solutions where $A\neq0$. We must further distinguish between whether $S=0$ or is non-trivial, however much of the analysis can be performed concurrently. 

Consider first the equation \eqref{A} for $A_{ij}$, which allows us to express the real part of $A^{*2}P$ in terms of $\Delta$ and $|A|$,
\be
\me^{-4 \Delta} \dd (\me^{4 \Delta} |A|^{2})= -(A^{*2}P+A^{2} P^{*})~.
\ee
Alternatively, from \eqref{APalg} we find
\begin{align}
\Re[A^{*2}P]&=-\frac{\me^{-4 \Delta}}{4} i_{J} (\star_{6} \GR-S \GI)~,\nonumber\\
\Im[A^{*2}P]&=\frac{\me^{-4 \Delta}}{4} i_{J}(\star_{6} \GI +S \GR)~,
\end{align}
where we have introduced the notation
\be
\GR\equiv \me^{2 \Delta}\Re[A \hat{G}^{*}]~,~~~\GI\equiv \me^{2 \Delta} \Im[A \hat{G}^{*}]~,~~~\hat{G}= \frac{\me^{2 \Delta}A}{|A|^{2}}(\GR-\ii \GI)~.
\ee
From the algebraic conditions \eqref{ImXintoG} and \eqref{BintoP} it is trivial to see that these three-forms have no $(3,0)$ nor $(0,3)$ part, $i_{\Re[\Omega]}\hat{G}=i_{\Im[\Omega]}\hat{G}=0$. Clearly since they are real they contain both a $(2,1)$-form and a $(1,2)$-form. Note that this differs with the $A=0$ case where the three-form $\hat{G}$ is either $(2,1)$ or $(1,2)$ but not both. Further the algebraic condition \eqref{dDeltaalg} implies 
\be
\dd S= \frac{\me^{-4 \Delta}}{2} i_{J} \GI~.
\ee
From \eqref{d8DeltaAnorm} and the scalar condition \eqref{SU(3)scalarcond} one can derive analogous expressions for $\dd \Delta$ and $\dd |A|$:
\begin{align}
\dd \Delta&= \frac{\me^{-4\Delta}}{8 |A|^2} i_{J}(\star_{6} \GR+S \GI)~,\\
\dd |A|&=-\frac{\me^{-4 \Delta}S}{2 |A|} i_{J} \GI~.
\end{align}
This fixes all the scalars of the theory in terms of contractions of the three-form with $J$.

Consistency of the three equations involving $J$, \eqref{dSJ}, \eqref{dJ} and \eqref{AdJ}, requires that the three-form flux satisfies
\be\label{Grelation}
\GI -\frac{1}{2} J \wedge i_{J} \GI=S \left( \star_{6} \GR -\frac{1}{2}J \wedge i_{J} \star_{6} \GR\right)~,
\ee
and then
\be\label{dJ1}
\me^{-4 \Delta} \dd (\me^{4 \Delta} J)=\me^{-4 \Delta}\left( \star_{6} \GR -\frac{1}{2}J \wedge i_{J} \star_{6} \GR\right)~.
\ee
Note that both $\dd J$ and $*\dd J$ are primitive and therefore equation \eqref{dJ1} implies the conformally balanced condition \eqref{balanced}.

The simplest way of solving \eqref{Grelation} is to set 
\be
\GI = S \star_{6} \GR
\ee
and this is in fact how our new solution satisfies this equation. However it is not clear that this is the most general solution.  What is clear is that when $S=0$ one has $\GI=0$ and many of the one-forms given above vanish. The $S=0$ case therefore results in a much simpler class of solutions to consider.

It remains to study the consistency of the three equations for the holomorphic three-form, namely equations, \eqref{AOmega}, \eqref{dAOmegab} and \eqref{dAOmega}. Using all the previously defined results it is simple to see that the three conditions all imply
\be\label{Omegadefining}
\me^{-6 \Delta} \dd (\me^{6 \Delta}\Omega)= \lb -2 \ii m K -\frac{\me^{-2 \Delta}}{|A|} \dd (\me^{2 \Delta}|A|)\rb \wedge \Omega~.
\ee
As previously stated we can see from the above that the base space is complex and that the complex structure is integrable. 

Similar to the $A=0$ case we absorb a conformal factor into the SU$(3)$ invariant forms, viz.
\be
j=m^{2} \me^{4 \Delta} J~,~~~\omega=m^{3}\me^{6 \Delta} \Omega~.
\ee
The equations then reduce to 
\begin{align}
\dd j &= \star_{6} \GR-\frac{1}{2} j \wedge i_{j} \star_{6} \GR~,\label{dj}\\
\dd \omega&=\lb -2\ii m K -\frac{\me^{-2 \Delta}}{|A|} \dd (\me^{2 \Delta}|A|)\rb\wedge \omega~,\label{domega}\\
\dd j \wedge j&=0~.\label{djj}
\end{align}
From \eqref{domega} we can compute the Ricci-form potential of the base space:
\be
\mathcal{P}= \sigma +\frac{\me^{-2 \Delta}}{|A|} \dd^{c} (\me^{2 \Delta}|A|),
\ee
where $\dd^{c}\equiv \ii (\bar{\partial}-\partial)= - i_{j}^{(1)} \dd$. The Ricci-form is given simply by $\rho = \dd \mathcal{P}$ and $\sigma$ is defined in \eqref{Kcomp}.

Since the base is not K\"ahler but only the weaker condition of being a balanced manifold some of the nice K\"ahler identities used in the previous section no longer hold. In particular the Ricci-form and the Levi-Civita curvature two-forms are not identical, see for example \cite{liu2014ricci}.  One can therefore define more than one scalar obtained by contracting the curvature two-form with the complex structure (i.e. the identity used to obtain the Ricci scalar in \eqref{KahleridentityR}). 
If we use that $i_{j} \dd K=-\frac{2}{m} \me^{-4 \Delta}$, as follows from \eqref{AdK}, we can compute an expression for the Chern Ricci scalar in terms of the warp factor and $|A|$:
\be
R_{C}=4 \me^{-4 \Delta} + i_{j}\dd \lb \frac{\me^{-2 \Delta}}{|A|} \dd^{c} (\me^{2 \Delta}|A|)\rb~.
\ee
For a balanced manifold the Chern Ricci scalar is related to the usual Ricci scalar by a torsion term
\be
R=2R_{C}-\frac{1}{2} |T|^2~,~~~~T=\dd^{c} j~.
\ee
The fluxes are determined via
\begin{align}
m F&=-2 S j - m \dd (\me^{4 \Delta} S K)~,\nonumber\\
G&=- A \me^{2 \Delta} K \wedge \dd K+ \frac{\me^{-2 \Delta}A}{|A|^{2}} (\GR-\ii \GI)~,\label{finalfluxes}\\
\gads&=-2m \me^{2 \Delta} A~, \nonumber
\end{align}
and the derivatives of the scalars are fixed via
\begin{align}
\dd S^2&=-\dd |A|^{2}= m^{2}S\, i_{j} \GI~,~~~\dd \Delta = \frac{m^{2}}{8|A|^{2}}\, i_{j}(\star_{6} \GR+ S \GI)~,~~~\nonumber\\
\Re[A^{*2}P]&= -\frac{m^{2}}{4} i_{j}(\star_{6} \GR -S \GI)~,~~~\Im[A^{*2} P]=\frac{m^{2}}{4} i_{j}i_{j}^{(1)}(S\star_{6} \GR - \GI)~.
\end{align}
Recall that all the equations of motion and Bianchi identities are satisfied provided that the Bianchi identity for $G$ is satisfied. This implies the unwieldy condition
\begin{align}
\me^{4 \Delta} |A|^{2}m^{2} \dd K \wedge \dd K =& \me^{4 \Delta} |A|^{2}\dd \dd^{c} \lb \frac{\me^{-4 \Delta}}{|A|^{2}}j \rb 
+ j \wedge \left[ -\frac{1+S^{2}}{2 |A|^{2}} \dd \dd^{c} S^{2}+\frac{2(1+S^{2})}{|A|^{2}}( \dd S^{2} \wedge \dd^{c} \Delta +\dd \Delta \wedge \dd^{c} S^{2})\right. \nonumber\\
&\left.+4 S^{2} \dd \dd^{c} \Delta -16 S^{2} \dd \Delta \wedge \dd^{c} \Delta+\frac{1- 6 S^{2}-3 S^{4}}{4 S^{2} |A|^{2}} \dd S^{2} \wedge \dd^{c} S^{2}\right]~.
\end{align}
Note that the last term vanishes if one imposes $\GI=S \star_{6}\GR$ or $S=0$. 

These are the general equations one needs to solve for a generic solution with all possible fluxes turned on.\footnote{It would be interesting to understand if there is a connection to the AdS$_5$ solutions of \cite{Gauntlett:2005ww} in an analogous way to the connection between the Sasaki--Einstein solutions and those of \cite{Kim:2005ez}. Since the solutions of \cite{Gauntlett:2005ww} admit an identity structure this seems somewhat fanciful.} This class of solutions allows for a very complicated brane construction, it would therefore be very interesting to find new solutions to this set of equations. We have presented a new solution in section \ref{Sec:Sols} derived using duality transformations and some other $\mathcal{N}=(0,4)$ solutions can be found in \cite{Lozano:2019emq} also found by using duality transformations.
 
The equations to solve are relatively imposing. To proceed it is useful to make an assumption and the most useful and less strict one is to set $\GI=S \star_{6}\GR$. Alternatively if one sets the stronger condition $S=0$ the conditions simplify further as we shall show now.

\subsection{Special case 2: $S=0$}\label{sec:S=0}

We set $S=0$ in the remainder of this section. Recall that this implies in addition that $|A|=1$. Clearly for $S=0$ the two-form $F$ vanishes and consequently there is no five-form flux.\footnote{Mirroring the previous footnote, it would be interesting to see if there is a connection with the AdS$_5$ classification in type IIB without five-form flux carried out in \cite{Couzens:2016iot} and this class of solutions.} Moreover it is easy to see that $\GI$ vanishes in this case, which in turn implies $\Im[A^{*2}P]=0$. 
We can rewrite \eqref{dj} as
\be\label{djS=0}
\me^{-4\Delta} \dd (\me^{4 \Delta} j)=\star_{6} \GR~,
\ee
whilst \eqref{domega} implies that the Ricci-form potential $\mathcal{P}$ is 
\be\label{PotentialS0}
\mathcal{P}=\sigma+2 \dd^{c} \Delta~,
\ee
Finally the Bianchi identity reduces to 
\be
\frac{1}{4} \dd \sigma \wedge \dd \sigma= \dd \dd^{c} (\me^{-4 \Delta} j)~.
\ee

An interesting ansatz to make is to set the warp factor to be constant. This is equivalent to imposing that the three-form flux $\star_{6}\GR$ is primitive. The equations reduce to the simpler set
\begin{align}
\rho&= \dd \sigma~,\\
\rho\wedge \rho&= 4 \dd \dd^{c} j~,\\
\dd j &=\star_{6}\GR~.
\end{align}
We shall present an example of a solution with this restriction in section \ref{Sec:Sols}, namely the classic AdS$_3\times S^3\times S^3 \times S^1$ solution.



\section{Examples of solutions}\label{Sec:Sols}

Having given the reduced torsion conditions for the three cases we shall exhibit how these can be solved for various solutions. We shall focus on the two cases where $A\neq0$ since the $A=0$ case has been studied, albeit not in full generality, in the literature already. For the $S=0$ case we shall show how the classic AdS$_3\times S^3\times S^3 \times S^1$ fits into the classification. Of course this solution preserves more than the $\mathcal{N}=(0,2)$ supersymmetry that we have imposed here and therefore we will restrict ourselves to an $\mathcal{N}=(0,2)$ subsector of the preserved supersymmetry. We will find that despite the warp factor and axio-dilaton being trivial, the flux $\GR$ is non-vanishing and is instead a primitive three-form on the base. Had we instead considered the AdS$_3\times S^3 \times \text{CY}_2$ with just three-form flux, we would have found that $\GR=0$. 

For the more generic case with both scalar bilinears non-vanishing we find a new solution by applying a beta deformation \cite{Lunin:2005jy} to the solution found in \cite{Gauntlett:2006ns,Donos:2008ug}. The solution we find has all fluxes, except axion, switched on. We shall make some brief comments on the field theory of the solution but we will not pursue this in more detail here since it can be understood from the beta deformed field theory of the seed solution.

\subsection{$S=0$ case and the $\ads_{3}\times S^3 \times S^3 \times S^1$ solution}

The near-horizon limit of the supergravity solution for the intersection of two stacks of D5-branes intersecting with the worldvolume of a stack of D1-branes was given in \cite{Cowdall:1998bu,Gauntlett:1998kc,Boonstra:1998yu}. The geometry is 
\be
\ads_{3}\times S^3_{+}\times S^{3}_{-}\times S^{1}
\ee
where we have compactified the real line one would get in the near-horizon. By suitable SL$(2,\Z)$ transformations one can exchange the RR-fluxes for NS-NS fluxes, or some combination of them. We shall take a solution with a constant parameter that allows us to interpolate between the two distinct endpoints. In particular the solution we will use is the following
\begin{align}
m^{2}\dd s^2&= \dd s^{2}(\ads_{3})+ \frac{R_{+}^{2}}{4 }(\sigma_{1+}^2+\sigma_{2+}^2+\sigma_{3+}^{2})+ \frac{R_{-}^{2}}{4}(\sigma_{1-}^{2}+\sigma_{2-}^{2}+\sigma_{3-}^{2})+ \dd z^{2}~,\nonumber\\
m^{2}G&=2\me^{\ii \varphi}\lb   \dd \vol(\ads_{3})+\frac{R_{+}^{2}}{8} \dd \vol(S^{3}_{+})+\frac{R_{-}^{2}}{8} \dd \vol(S^{3}_{-})\rb~,
\end{align}
where $\sigma_{i \pm}$ are the SU$(2)$ invariant Maurer-Cartan one-forms, $R_{\pm}$ are the radii of the two three-spheres satisfying $R_{+}^{2}R_{-}^{2}=R_{-}^{2}+R_{+}^{2}$, the radius of AdS$_3$ is given by $m^2=R_{+}^{2}+R_{-}^{2}$ and $\varphi$ is a real constant. We have defined the volume of the three-spheres to be $\dd \vol(S^{3}_{\pm})= \sigma_{1\pm}\wedge \sigma_{2\pm} \wedge \sigma_{3\pm}$. All other fluxes are trivial. We shall focus only on an $\mathcal{N}=(0,2)$ subsector of the total preserved supersymmetry, the choice of which amongst the four quasi-Majorana spinors satisfying \eqref{quasi-maj} is arbitrary. It is easy to find the Killing spinors solving \eqref{PKSE}-\eqref{dKSE} on the internal manifold:
\be
\xi=\frac{1}{2}\left\{\me^{\ii \varphi}\bar{a},-\me^{\ii \varphi} \bar{b},\me^{\ii \varphi}  \frac{\ii R_{+}+R_{-}}{R_{+}R_{-}} \bar{a},-\me^{\ii \varphi}\frac{R_{-}-\ii R_{+}}{R_{+}R_{-}} \bar{b},\frac{\ii R_{+}+R_{-}}{R_{+}R_{-}}b,\frac{R_{-}-\ii R_{+}}{R_{+}R_{-}} a,b,a\right\}~,
\ee
with $a$ and $b$ arbitrary complex constants. Moreover it is not too difficult to see that the spinor satisfies the quasi-Majorana condition \eqref{quasi-maj} as required. In the following we shall work with the SU$(3)$ structure given by the two Killing spinors obtained by setting $a=0$, $b=1$ for the first and $a=0$, $b=\ii$ for the second. We reiterate that this is an arbitrary choice and any other combination would also give an SU$(3)$ structure satisfying the conditions presented above. Then the scalar bilinears are
\be
A= -\me^{\ii \varphi}~,~~~S_{ij}= \delta_{ij}~.
\ee
The Killing vector is 
\be 
K=\frac{1}{2m}(\sigma_{1-}-\sigma_{3+})~,
\ee
whilst the two-form $j$ is
\be
m^{2} j=\frac{R_{+}^{2}}{4}\sigma_{1+}\wedge \sigma_{2+}-\frac{R_{-}^{2}}{4} \sigma_{2-}\wedge \sigma_{3-}-\frac{1}{2}\lb \frac{R_{1}}{R_{2}} \sigma_{3+}+\frac{R_{2}}{R_{1}} \sigma_{1-}\rb \wedge \dd z~.
\ee
Finally the three-form $\GR$ is given by
\vspace{7mm}
\begin{align}
\GR=&\frac{1}{4 m^{2}R_{-}^{3}}[ R_{+}^{2}R_{-} \sigma_{1+}\wedge \sigma_{2+}\wedge \sigma_{3+}+R_{+}^{2}R_{-}(R_{-}^{2}-1) \sigma_{1+}\wedge \sigma_{2+} \wedge \sigma_{1-}\nonumber\\
&+ R_{-}^{3} \sigma_{3+}\wedge \sigma_{2-}\wedge \sigma_{3-} +R_{-}^3(R_{-}^{2}-1) \sigma_{1-}\wedge \sigma_{2-}\wedge \sigma_{3-}]~,
\end{align}
which can be seen to be primitive. It is now simple to check that the reduced torsion conditions of section \ref{sec:S=0} are all satisfied. 


\subsection{A representative solution of the general case}

In this section we shall present a representative example of the generic class of solutions. This solution was obtained by taking the Baryonic twist solution for $Y^{p,q}$ found in \cite{Gauntlett:2006ns} and further elucidated in \cite{Couzens:2017nnr} and performing a beta deformation \cite{Lunin:2005jy} along the two directions of the two-torus in the geometry. We use the conventions for the solution presented in \cite{Couzens:2017nnr}. The final solution includes all fluxes but for a non-trivial axion, though this may be added by using the SL$(2,\Z)$ symmetry. We shall present only the final solution since performing a beta deformation on a two-torus is relatively straightforward. The final solution is 
\begin{align}
m^{2} \dd s^{2}(X_{7})&=\frac{1}{4}(\dd \chi+\sigma)^{2}+\me^{-4 \Delta}\left[\frac{\sqrt{1+a \beta^{2}  x}}{4 x} \left( \frac{\dd x^{2}}{x^{2}U(x)}+U(x) D \psi^{2}+ \dd s^{2}(S^{2})_{\theta,\phi}\right)+ \dd x_{1}^{2} +\dd x_{2}^{2}\right]~,\nonumber\\
\me^{-4 \Delta}&=\frac{a x}{\sqrt{1+a \beta^{2} x}}~,~~U(x)=1-a(x-1)^{2}~,\\
\sigma&=a(x-1)  D\psi +2 \dd \psi~,\nonumber
\end{align}
where we have defined the shorthand $D\psi=\dd \psi+ \cos\theta \dd \phi$. The fluxes are given by
\begin{align}
m^{2}G&= - \frac{2\ii  \beta}{(1+ a \beta^{2} x)^{\frac{1}{4}}}\dd \vol(\ads_{3})-A\me^{2 \Delta} K\wedge \dd K+ \frac{\me^{-2 \Delta}A}{|A|^{2}}\GR ~,~~\nonumber\\
\me^{-2 \Delta}\GR&= \frac{\ii a \beta^{2} \sqrt{a x}}{(1+ a \beta^{2} x)^{\frac{9}{4}}} \dd x \wedge \dd \vol(T^2) + \frac{ a \beta^{2} \sqrt{ax} U(x)}{4 (1+ a \beta^{2} x)^{\frac{3}{4}}}D \psi \wedge \dd \vol(S^{2})~,~~\nonumber\\
\me^{-2 \Phi}&= 1+ a \beta^{2} x~,\\
A&=\frac{\ii \beta \sqrt{ax}}{ \sqrt{1+ a \beta^{2} x}}~,\nonumber\\
m F&=-\frac{1}{2 a x^{2} }\dd x \wedge D \chi +\frac{1-x}{2 x^{2}} \dd x \wedge D \psi + \frac{2}{(1+ a \beta^{2} x)} +\frac{1}{2} \dd \vol(S^{2})~.\nonumber
\end{align}
We have written the metric and fluxes in terms of the classification and it is easy to see that the conditions are satisfied identically. 

From the origin of the solution it is easy to see that the metric is regular by construction and therefore we shall not perform any regularity analysis. Furthermore by construction the central charge will be $\beta$ independent and therefore its result precisely agrees with the value of the central charge of the seed solution. 
It is interesting to note that the field theories on the two sides have quite different brane constructions. The original seed theory consists of wrapping D3-branes on a two-torus with a Baryonic twist whilst this new solution is a complicated intersection of D3-branes, (p,q) 5-branes and fundamental strings with a non-trivial dilaton. It would be interesting to better understand the dual field theory in this case.



\section{Concluding remarks}\label{Sec:Conclude}

In this paper we have classified all AdS$_3$ solutions of type IIB supergravity with an AdS$_3$ factor, arbitrary fluxes and an SU$(3)$ structure. We have seen that there are three distinct cases to consider depending on the values of particular scalar bilinears. 

The first case recovers the master equation derived in \cite{Passias:2019rga} which is a generalization of the master equations appearing in \cite{Kim:2005ez, Donos:2008ug, Couzens:2017nnr}. The internal manifold is a $U(1)$ fibration over a conformally K\"ahler base space which supports a holomorphically varying axio-dilaton, a particular choice of three-form flux (primitive and of $(2,1)$ type), and in addition the geometry supports five-form flux. Only this case, of the three discussed in this paper, allows for a canonical F-theoretic interpretation. Duality to eleven-dimensional supergravity along the AdS$_3$ factor (see \cite{Couzens:2017nnr} for further details of the duality prescribed here) shows that this class of solutions are dual to the AdS$_2$ class discussed in \cite{Donos:2008ug} when the K\"ahler base is taken to be an elliptically fibered manifold satisfying the eleven-dimensional master equation in \cite{Donos:2008ug}. 
Of the three cases studied in the paper it is this first class that looks most promising for extending the geometric dual of c-extremization. 

The second special class of supersymmetric solutionsconsist of $U(1)$ fibrations over a complex base, however unlike in the previous case this is not generically K\"ahler, but instead admit a conformally balanced metric. In this case the geometry is supported by three-form flux and admits a non-trivial axio-dilaton, albeit not a holomorphically varying one. It is interesting to note that it is necessary to have both 1-branes and 5-branes in such solutions, with the classic D1-D5 (or their SL$(2,\Z)$ relatives) solutions examples of geometries in this class. 

The final class of solutions allows for the inclusion of all fluxes. As before the internal manifold is a $U(1)$ fibration over a complex conformally balanced manifold, and again is not generically K\"ahler. We have provided an example of a solution within this class which was obtained by performing a beta deformation to a known solution discussed in \cite{Gauntlett:2006ns}. Regularity of the solution follows from regularity of the seed solution which was performed in \cite{Gauntlett:2006ns} and expanded on in \cite{Couzens:2017nnr}. Moreover the central charge in the beta deformed solution is known, by general arguments of beta deformation and T-dualities to be independent of the deformation parameter and therefore it is the same as the result presented in \cite{Couzens:2017nnr}. It would still be interesting to study the field theory in more detail. 

We have presented the necessary and sufficient conditions for a bosonic AdS$_3$ geometry with SU$(3)$ structure and generic fluxes. It would be interesting to find the geometric dual of c-extremization for this more general class of solutions. The first class seems the most promising of the three discussed here due to its similarity with the class of solutions considered in \cite{Couzens:2018wnk, Gauntlett:2018dpc,Gauntlett:2019pqg}.


\section*{Acknowledgments}

It is a pleasure to thank Thomas Grimm, Huibert het Lam, Niall Macpherson, Dario Martelli, Kilian Mayer, and Sakura Sch\"afer-Nameki for useful discussions at various stages of this project. In addition, thanks go to Pyry Kuusela for collaboration at an earlier stage of this project.
The author acknowledges the support of the Netherlands Organization for Scientifc Research (NWO) under the VICI grant 680-47-602.

\appendix




\section{Imposing an SU$(3)$ G-structure}\label{app:SU(3)}

In general a Dirac spinor in seven dimensions admits an SU$(3)$ structure which defines a vector $v$ and a transverse foliation admitting a canonical SU$(3)$ structure, $(J, \Omega)$. We are interested in imposing $\mathcal{N}=(0,2)$ supersymmetry and therefore we assume the existence of two independent nowhere vanishing Dirac spinors $\xi_{i}$. Clearly each Killing spinor defines its own SU$(3)$ G-structure From two such spinors it is possible to obtain three possible G-structures depending on the intersection of the SU$(3)$ structures: SU$(3)$, SU$(2)$ and an identity structure. In this appendix we shall focus on the case where the two SU$(3)$ structures intersect, giving an SU$(3)$ structure. We will find that the existence of such an SU$(3)$ structure implies the condition on the scalar bilinears given in the main text \eqref{SU(3)scalarcond}.
 
Since a Dirac spinor in seven dimensions defines a vector $v$ we can define a unit norm vector $e^7= \frac{v}{|v|}$. Define the projectors $P_{\pm}=\frac{1}{2}(1+\pm \gamma^{7})$ and compute the projections of $\xi$, $\xi_{\pm}=\xi$. The six-dimensional space transverse to the vector $e^7$ then admits positive and negative chirality spinors under the action of $\gamma^{7}$. The projections $\xi_{\pm}$ can then be associated with positive (negative) chirality spinors in six dimensions respectively. Each spinor then defines an SU$(3)$ structure since SU$(3)$ is the stabilizer of a six-dimensional spinor of definite chirality. One can argue (though we suppress the details here) that the two spinors are related via
\be
\xi_{+}= \lambda \xi_{-}^{c}
\ee
with $\lambda$ a complex function and ${}^{c}$ charge conjugation.

This allows us to decompose an arbitrary 7d Dirac spinor as
\be
\xi =a \chi + b \chi^{c}~,~~~ \gamma^7 \chi=\chi~, ~~~ \bar{\chi}\chi=1~.
\ee
With the second spinor we may run the same argument. Since we are looking for an SU$(3)$ structure it must be that the two vectors are aligned and therefore the unit norm vectors differ only by an irrelevant sign. We can now expand both spinors in terms of the $\chi$ above:\footnote{If we wanted to look at an SU$(2)$ structure one should now introduce a second spinor $\hat{\chi}$ such that $\xi_{2}= c \chi+ d \chi^{c}+ \hat{c} \hat{\chi}+\hat{d} \hat{\chi}^{c}$. It is simple to see that imposing an SU$(3)$ structure would set $\hat{c}=\hat{d}=0$ as expected.}
\be\label{xisdecomp}
\xi_{1}= a \chi+b \chi^{c}~,~~~\xi_{2}=c \chi+ d \chi^{c}~.
\ee
Since $\chi$ is chiral it satisfies $\bar{\chi}^c\chi=0$ and $\bar{\chi}^{c} \gamma_{(1)} \chi=0$. We may now compute the various bilinears in terms of this basis. We find
\begin{align}
S_{11}&=|a|^2+|b|^2=1~,~~S_{22}= |c|^2+|d|^2=1~,~~S_{12}=a^* c+ b d^*=\ii S~,~~\nonumber\\
 A_{11}&= 2 a b= 2 c d= A_{22}\equiv A~,~~A_{12}= a d+b c~.
\end{align}
We may solve these conditions via the real functions $\Theta$ and $\varphi$ as
\begin{align}\label{xisdecompfunctions}
a=\me^{\ii \varphi}\sin \frac{\Theta}{2} ~,~~ b=\cos \frac{\Theta}{2}~,~~c=-\ii\me^{\ii \varphi} \sin \frac{\Theta}{2}~,~~d= \ii  \cos \frac{\Theta}{2}~,
\end{align}
which implies that the scalars are given by
\be
S=\cos \Theta~,~~ A= \me^{\ii \varphi} \sin \Theta~.
\ee
From here it is then trivial to check that imposing an SU$(3)$ structure implies equation \eqref{SU(3)scalarcond}:
\be
1=|A|^2 + S^2~.
\ee
The one-form bilinears in this basis are easily computed to be
\be
\Im[K_{12}]\equiv K=-e^7~,~~K_{11}=K_{22} = S K~,~~\Re[K_{12}]=0~,~~B_{12}= \ii A K~.
\ee
The two-form bilinears are given by
\begin{align}
U_{12}=J~,~~U_{11}=U_{22}=-\ii S J~,~~V_{12}= A J~,
\end{align}
whilst the three-form bilinears are
\begin{align}
X_{11}&=-\ii K \wedge J - \ii |A| \Re \Omega~,~~X_{22}=-\ii K \wedge J + \ii |A| \Re\Omega~,~~X_{12}= S K\wedge J -\ii |A| \Im \Omega~,\nonumber\\
Y_{11}&=-\ii A K\wedge J - \ii \me^{\ii \varphi}( \Re \Omega -\ii S \Im \Omega)~,~~Y_{22}= -\ii K \wedge J + \ii \me^{\ii \varphi} ( \Re \Omega -\ii S \Im \Omega)~,\nonumber\\
Y_{12}&=\me^{\ii \varphi} ( S \Re \Omega - \ii \Im \Omega)~.
\end{align}

\subsubsection*{Special limits}

Before we end this section we shall look at the special limits of the function $\Theta$. For $\Theta=0$ we have that $A$ vanishes and $S=1$. From equation \eqref{xisdecomp} and \eqref{xisdecompfunctions} we see that this imposes $\xi_{1}= \ii \xi_{2}$. If we insert this into the Killing spinor equations \eqref{PKSE}-\eqref{dKSE} we see that the $G^{(3)}$ dependent part of the supersymmetry equations decouples from the remainder. One is left with the Killing spinor equations of \cite{Couzens:2017nnr} coupled with the $G^{(3)}$ dependent part of \cite{Donos:2008ug}. It is then clear that this is precisely the class of solutions considered in section 5 of \cite{Passias:2019rga} which generalises the \emph{master equations} of \cite{Kim:2005ez, Donos:2008ug, Couzens:2017nnr} in various directions. 


The other special limit is $\Theta= \frac{\pi}{2}$. In this case $S$ vanishes and $A$ is just a phase. From \eqref{xisdecomp} and \eqref{xisdecompfunctions} we see that the spinors $\xi_{i}$ become quasi Majorana
\be\label{quasi-maj}
\xi_{i}^{c}= - \me^{-\ii \varphi} \xi_{i}~,
\ee
with the phase correlated with the phase of the bilinear $A$ and therefore also $\gads$. As we saw in the main text in this limit the five-form flux is switched off and solutions only contain three-form fluxes. Moreover it is necessary to include both non-trivial $\gads$ and $G$ flux, i.e. both magnetic and electric components of three-form flux in ten dimensions. It is in this class that the well-known D1-D5 solutions\footnote{In fact any SL$(2,\mathbb{Z})$ duality frame of these solutions.} AdS$_3\times S^3 \times S^3\times S^1$ and AdS$_3 \times S^3 \times CY_2$ (or quotients thereof) can be embedded when studying an $\mathcal{N}=(0,2)$ subset of their full supersymmetry.

\section{Useful identities}\label{identities}

In this final section we gather some useful identities that we have employed in the various sections for simplifying the conditions on the base. Since neither $(3,0)$- nor $(0,3)$-forms played any role in this paper we shall assume that all three-forms in this section are of $(2,1)$ or $(1,2)$ type only. Let $Z$ be such a three-form, then
\begin{align}
i_{J} \star_{6} Z&= \star_{6} (J \wedge Z)~,\nonumber\\
i_{J}Z&=-\star_{6}(J \wedge \star_{6} Z)~,\nonumber\\
\star_{6} Z&=i^{(1)}_{J}\lb Z -J\wedge ( i_{J} Z)\rb~,\nonumber\\
i_{J}^{(1)} (i_{J} Z)&=-i_{J}\star_{6}Z~.
\end{align}
Moreover for a one-form $z$ on the base one has
\begin{align}
\star_{6} \lb z \wedge\frac{1}{2} J^{2}\rb&= -i_{J}^{(1)} z~,\nonumber\\
\star_{6} z\wedge J&= -J\wedge (i_{J}^{(1)}z)~.
\end{align}
More useful identities can be found in the appendix of \cite{Passias:2019rga}.





\bibliographystyle{JHEP}

\bibliography{ADSCFT}

\providecommand{\href}[2]{#2}\begingroup\raggedright\begin{thebibliography}{10}

\bibitem{Benini:2012cz}
F.~Benini and N.~Bobev, \emph{{Exact two-dimensional superconformal R-symmetry
  and c-extremization}},
  \href{http://dx.doi.org/10.1103/PhysRevLett.110.061601}{\emph{Phys. Rev.
  Lett.} {\bf 110} (2013) 061601}, [\href{https://arxiv.org/abs/1211.4030}{{\tt
  1211.4030}}].

\bibitem{Benini:2013cda}
F.~Benini and N.~Bobev, \emph{{Two-dimensional SCFTs from wrapped branes and
  c-extremization}},
  \href{http://dx.doi.org/10.1007/JHEP06(2013)005}{\emph{JHEP} {\bf 06} (2013)
  005}, [\href{https://arxiv.org/abs/1302.4451}{{\tt 1302.4451}}].

\bibitem{Couzens:2018wnk}
C.~Couzens, J.~P. Gauntlett, D.~Martelli and J.~Sparks, \emph{{A geometric dual
  of $c$-extremization}},
  \href{http://dx.doi.org/10.1007/JHEP01(2019)212}{\emph{JHEP} {\bf 01} (2019)
  212}, [\href{https://arxiv.org/abs/1810.11026}{{\tt 1810.11026}}].

\bibitem{Gauntlett:2018dpc}
J.~P. Gauntlett, D.~Martelli and J.~Sparks, \emph{{Toric geometry and the dual
  of $c$-extremization}},
  \href{http://dx.doi.org/10.1007/JHEP01(2019)204}{\emph{JHEP} {\bf 01} (2019)
  204}, [\href{https://arxiv.org/abs/1812.05597}{{\tt 1812.05597}}].

\bibitem{Gauntlett:2019pqg}
J.~P. Gauntlett, D.~Martelli and J.~Sparks, \emph{{Fibred GK geometry and
  supersymmetric $AdS$ solutions}},
  \href{https://arxiv.org/abs/1910.08078}{{\tt 1910.08078}}.

\bibitem{Hosseini:2019use}
S.~M. Hosseini and A.~Zaffaroni, \emph{{Proving the equivalence of
  $c$-extremization and its gravitational dual for all toric quivers}},
  \href{http://dx.doi.org/10.1007/JHEP03(2019)108}{\emph{JHEP} {\bf 03} (2019)
  108}, [\href{https://arxiv.org/abs/1901.05977}{{\tt 1901.05977}}].

\bibitem{Kim:2005ez}
N.~Kim, \emph{{AdS$_3$ solutions of IIB supergravity from D3-branes}},
  \href{http://dx.doi.org/10.1088/1126-6708/2006/01/094}{\emph{JHEP} {\bf 01}
  (2006) 094}, [\href{https://arxiv.org/abs/hep-th/0511029}{{\tt
  hep-th/0511029}}].

\bibitem{Gauntlett:2007ts}
J.~P. Gauntlett and N.~Kim, \emph{{Geometries with Killing Spinors and
  Supersymmetric AdS Solutions}},
  \href{http://dx.doi.org/10.1007/s00220-008-0575-5}{\emph{Commun. Math. Phys.}
  {\bf 284} (2008) 897--918}, [\href{https://arxiv.org/abs/0710.2590}{{\tt
  0710.2590}}].

\bibitem{Donos:2008ug}
A.~Donos, J.~P. Gauntlett and N.~Kim, \emph{{AdS Solutions Through
  Transgression}},
  \href{http://dx.doi.org/10.1088/1126-6708/2008/09/021}{\emph{JHEP} {\bf 09}
  (2008) 021}, [\href{https://arxiv.org/abs/0807.4375}{{\tt 0807.4375}}].

\bibitem{Couzens:2017way}
C.~Couzens, C.~Lawrie, D.~Martelli, S.~Schafer-Nameki and J.-M. Wong,
  \emph{{F-theory and AdS$_{3}$/CFT$_{2}$}},
  \href{http://dx.doi.org/10.1007/JHEP08(2017)043}{\emph{JHEP} {\bf 08} (2017)
  043}, [\href{https://arxiv.org/abs/1705.04679}{{\tt 1705.04679}}].

\bibitem{Eberhardt:2017uup}
L.~Eberhardt, \emph{{Supersymmetric $\mathrm{AdS}_3$ supergravity backgrounds
  and holography}},  \href{https://arxiv.org/abs/1710.09826}{{\tt 1710.09826}}.

\bibitem{Couzens:2017nnr}
C.~Couzens, D.~Martelli and S.~Schafer-Nameki, \emph{{F-theory and
  AdS$_{3}$/CFT$_{2}$ (2, 0)}},
  \href{http://dx.doi.org/10.1007/JHEP06(2018)008}{\emph{JHEP} {\bf 06} (2018)
  008}, [\href{https://arxiv.org/abs/1712.07631}{{\tt 1712.07631}}].

\bibitem{Passias:2019rga}
A.~Passias and D.~Prins, \emph{{On $\mathcal{N}=1$ AdS$_3$ solutions of Type
  IIB}},  \href{https://arxiv.org/abs/1910.06326}{{\tt 1910.06326}}.

\bibitem{Gauntlett:2006ns}
J.~P. Gauntlett, N.~Kim and D.~Waldram, \emph{{Supersymmetric AdS$_3$, AdS$_2$
  and Bubble Solutions}},
  \href{http://dx.doi.org/10.1088/1126-6708/2007/04/005}{\emph{JHEP} {\bf 04}
  (2007) 005}, [\href{https://arxiv.org/abs/hep-th/0612253}{{\tt
  hep-th/0612253}}].

\bibitem{Gauntlett:2006af}
J.~P. Gauntlett, O.~A.~P. Mac~Conamhna, T.~Mateos and D.~Waldram,
  \emph{{Supersymmetric AdS(3) solutions of type IIB supergravity}},
  \href{http://dx.doi.org/10.1103/PhysRevLett.97.171601}{\emph{Phys. Rev.
  Lett.} {\bf 97} (2006) 171601},
  [\href{https://arxiv.org/abs/hep-th/0606221}{{\tt hep-th/0606221}}].

\bibitem{Benini:2015bwz}
F.~Benini, N.~Bobev and P.~M. Crichigno, \emph{{Two-dimensional SCFTs from
  D3-branes}}, \href{http://dx.doi.org/10.1007/JHEP07(2016)020}{\emph{JHEP}
  {\bf 07} (2016) 020}, [\href{https://arxiv.org/abs/1511.09462}{{\tt
  1511.09462}}].

\bibitem{toappear}
C.~Couzens, H.~het Lam and K.~Mayer, \emph{{To appear}}, .

\bibitem{Jeong:2014iva}
J.~Jeong, E.~Ó~Colgáin and K.~Yoshida, \emph{{SUSY properties of warped
  $AdS_3$}}, \href{http://dx.doi.org/10.1007/JHEP06(2014)036}{\emph{JHEP} {\bf
  06} (2014) 036}, [\href{https://arxiv.org/abs/1402.3807}{{\tt 1402.3807}}].

\bibitem{Lozano:2019ywa}
Y.~Lozano, N.~T. Macpherson, C.~Nunez and A.~Ramirez, \emph{{AdS$_3$ solutions
  in massive IIA, defect CFTs and T-duality}},
  \href{https://arxiv.org/abs/1909.11669}{{\tt 1909.11669}}.

\bibitem{Lozano:2019zvg}
Y.~Lozano, N.~T. Macpherson, C.~Nunez and A.~Ramirez, \emph{{Two dimensional
  ${\cal N}=(0,4)$ quivers dual to AdS$_3$ solutions in massive IIA}},
  \href{https://arxiv.org/abs/1909.10510}{{\tt 1909.10510}}.

\bibitem{Lozano:2019emq}
Y.~Lozano, N.~T. Macpherson, C.~Nunez and A.~Ramirez, \emph{{AdS$_3$ solutions
  in Massive IIA with small $\mathcal{N}=(4,0)$ supersymmetry}},
  \href{https://arxiv.org/abs/1908.09851}{{\tt 1908.09851}}.

\bibitem{Lozano:2019jza}
Y.~Lozano, N.~T. Macpherson, C.~Nunez and A.~Ramirez, \emph{{1/4 BPS
  AdS$_3$/CFT$_2$}},  \href{https://arxiv.org/abs/1909.09636}{{\tt
  1909.09636}}.

\bibitem{Lozano:2015bra}
Y.~Lozano, N.~T. Macpherson, J.~Montero and E.~O. Colg\'ain, \emph{{New AdS$_3
  \times S^2$ T-duals with $ \mathcal{N}=\left(0,4\right) $ supersymmetry}},
  \href{http://dx.doi.org/10.1007/JHEP08(2015)121}{\emph{JHEP} {\bf 08} (2015)
  121}, [\href{https://arxiv.org/abs/1507.02659}{{\tt 1507.02659}}].

\bibitem{Macpherson:2018mif}
N.~T. Macpherson, \emph{{Type II solutions on AdS$_{3} \times$ S$^{3} \times$
  S$^{3}$ with large superconformal symmetry}},
  \href{http://dx.doi.org/10.1007/JHEP05(2019)089}{\emph{JHEP} {\bf 05} (2019)
  089}, [\href{https://arxiv.org/abs/1812.10172}{{\tt 1812.10172}}].

\bibitem{Dibitetto:2018ftj}
G.~Dibitetto, G.~Lo~Monaco, A.~Passias, N.~Petri and A.~Tomasiello,
  \emph{{AdS$_3$ Solutions with Exceptional Supersymmetry}},
  \href{http://dx.doi.org/10.1002/prop.201800060}{\emph{Fortsch. Phys.} {\bf
  66} (2018) 1800060}, [\href{https://arxiv.org/abs/1807.06602}{{\tt
  1807.06602}}].

\bibitem{Beck:2015gqa}
S.~W. Beck, J.~B. Gutowski and G.~Papadopoulos, \emph{{Geometry and
  supersymmetry of heterotic warped flux AdS backgrounds}},
  \href{http://dx.doi.org/10.1007/JHEP07(2015)152}{\emph{JHEP} {\bf 07} (2015)
  152}, [\href{https://arxiv.org/abs/1505.01693}{{\tt 1505.01693}}].

\bibitem{Martelli:2003ki}
D.~Martelli and J.~Sparks, \emph{{G structures, fluxes and calibrations in M
  theory}}, \href{http://dx.doi.org/10.1103/PhysRevD.68.085014}{\emph{Phys.
  Rev.} {\bf D68} (2003) 085014},
  [\href{https://arxiv.org/abs/hep-th/0306225}{{\tt hep-th/0306225}}].

\bibitem{Colgain:2010wb}
E.~O~Colgain, J.-B. Wu and H.~Yavartanoo, \emph{{Supersymmetric AdS$_3 \times
  S^2$ M-theory geometries with fluxes}},
  \href{http://dx.doi.org/10.1007/JHEP08(2010)114}{\emph{JHEP} {\bf 08} (2010)
  114}, [\href{https://arxiv.org/abs/1005.4527}{{\tt 1005.4527}}].

\bibitem{Gauntlett:2002sc}
J.~P. Gauntlett, D.~Martelli, S.~Pakis and D.~Waldram, \emph{{G structures and
  wrapped NS5-branes}},
  \href{http://dx.doi.org/10.1007/s00220-004-1066-y}{\emph{Commun. Math. Phys.}
  {\bf 247} (2004) 421--445}, [\href{https://arxiv.org/abs/hep-th/0205050}{{\tt
  hep-th/0205050}}].

\bibitem{Gauntlett:2005ww}
J.~P. Gauntlett, D.~Martelli, J.~Sparks and D.~Waldram, \emph{{Supersymmetric
  AdS$_5$ solutions of type IIB supergravity}},
  \href{http://dx.doi.org/10.1088/0264-9381/23/14/009}{\emph{Class. Quant.
  Grav.} {\bf 23} (2006) 4693--4718},
  [\href{https://arxiv.org/abs/hep-th/0510125}{{\tt hep-th/0510125}}].

\bibitem{Kuusela:2019iok}
P.~Kuusela, \emph{{"GammaMaP" - A Mathematica Package for Clifford Algebras,
  Gamma Matrices and Spinors}},  \href{https://arxiv.org/abs/1905.00429}{{\tt
  1905.00429}}.

\bibitem{Gauntlett:2019roi}
J.~P. Gauntlett, D.~Martelli and J.~Sparks, \emph{{Toric geometry and the dual
  of ${\cal I}$-extremization}},
  \href{http://dx.doi.org/10.1007/JHEP06(2019)140}{\emph{JHEP} {\bf 06} (2019)
  140}, [\href{https://arxiv.org/abs/1904.04282}{{\tt 1904.04282}}].

\bibitem{Hosseini:2019ddy}
S.~M. Hosseini and A.~Zaffaroni, \emph{{Geometry of $\mathcal{I}$-extremization
  and black holes microstates}},
  \href{http://dx.doi.org/10.1007/JHEP07(2019)174}{\emph{JHEP} {\bf 07} (2019)
  174}, [\href{https://arxiv.org/abs/1904.04269}{{\tt 1904.04269}}].

\bibitem{Kim:2019umc}
H.~Kim and N.~Kim, \emph{{Black holes with baryonic charge and
  $\mathcal{I}$-extremization}},  \href{https://arxiv.org/abs/1904.05344}{{\tt
  1904.05344}}.

\bibitem{Lunin:2005jy}
O.~Lunin and J.~M. Maldacena, \emph{{Deforming field theories with U(1) x U(1)
  global symmetry and their gravity duals}},
  \href{http://dx.doi.org/10.1088/1126-6708/2005/05/033}{\emph{JHEP} {\bf 05}
  (2005) 033}, [\href{https://arxiv.org/abs/hep-th/0502086}{{\tt
  hep-th/0502086}}].

\bibitem{liu2014ricci}
K.~Liu and X.~Yang, \emph{Ricci curvatures on hermitian manifolds},  2014.

\bibitem{Couzens:2016iot}
C.~Couzens, \emph{{Supersymmetric AdS$_{5}$ solutions of type IIB supergravity
  without D3 branes}},
  \href{http://dx.doi.org/10.1007/JHEP01(2017)041}{\emph{JHEP} {\bf 01} (2017)
  041}, [\href{https://arxiv.org/abs/1609.05039}{{\tt 1609.05039}}].

\bibitem{Cowdall:1998bu}
P.~M. Cowdall and P.~K. Townsend, \emph{{Gauged supergravity vacua from
  intersecting branes}}, \href{http://dx.doi.org/10.1016/S0370-2693(98)00768-0,
  10.1016/S0370-2693(98)00445-6}{\emph{Phys. Lett.} {\bf B429} (1998)
  281--288}, [\href{https://arxiv.org/abs/hep-th/9801165}{{\tt
  hep-th/9801165}}].

\bibitem{Gauntlett:1998kc}
J.~P. Gauntlett, R.~C. Myers and P.~K. Townsend, \emph{{Supersymmetry of
  rotating branes}},
  \href{http://dx.doi.org/10.1103/PhysRevD.59.025001}{\emph{Phys. Rev.} {\bf
  D59} (1998) 025001}, [\href{https://arxiv.org/abs/hep-th/9809065}{{\tt
  hep-th/9809065}}].

\bibitem{Boonstra:1998yu}
H.~J. Boonstra, B.~Peeters and K.~Skenderis, \emph{{Brane intersections,
  anti-de Sitter space-times and dual superconformal theories}},
  \href{http://dx.doi.org/10.1016/S0550-3213(98)00512-4}{\emph{Nucl. Phys.}
  {\bf B533} (1998) 127--162},
  [\href{https://arxiv.org/abs/hep-th/9803231}{{\tt hep-th/9803231}}].

\end{thebibliography}\endgroup

\end{document}